# High-resolution spectroscopy of single nuclear spins via sequential weak measurements

Matthias Pfender[1,*], Ping Wang[2,3,*], Hitoshi Sumiya[4], Shinobu Onoda[5], Wen Yang[2], Durga Bhaktavatsala Rao Dasari[1,7], Philipp Neumann[1], Xin-Yu Pan[8], Junichi Isoya[6], Ren-Bao Liu[3†], J. Wrachtrup[1,7†]

1. 3rd Institute of Physics, Research Center SCoPE and IQST, University of Stuttgart, 70569 Stuttgart, Germany
2. Beijing Computational Science Research Center, Beijing 100193, China
3. Department of Physics & Centre for Quantum Coherence, The Chinese University of Hong Kong, Shatin, New Territories, Hong Kong, China
4. Sumitomo Electric Industries Ltd., Itami 664-0016, Japan
5. Takasaki Advanced Radiation Research Institute, National Institutes for Quantum and Radiological Science and Technology, Takasaki 370-1292, Japan
6. Research Center for Knowledge Communities, University of Tsukuba, Tsukuba 305-8550, Japan
7. Max Planck Institute for Solid State Research, Stuttgart, Germany
8. Institute of Physics, Chinese Academy of Sciences, Beijing 100190, China

* These authors contributed equally

† Correspondence should be addressed to J.W. (wrachtrup@physik.uni-stuttgart.de) or R.B.L. (rbliu@cuhk.edu.hk).

**Quantum sensors have recently achieved to detect the magnetic moment of few or single nuclear spins and measure their magnetic resonance (NMR) signal. However, the spectral resolution, a key feature of NMR, has been limited by relaxation of the sensor to a few kHz at room temperature. The spectral resolution of NMR signals from single nuclear spins can be improved by, e.g., using quantum memories, however at the expense of sensitivity. Classical signals on the other hand can be measured with exceptional spectral resolution by using continuous measurement techniques, without compromising sensitivity. To apply these techniques to single-spin NMR, it is critical to overcome the impact of back action inherent of quantum measurements. Here we report sequential weak measurements on a single $^{13}C$ nuclear spin. The back-action of repetitive weak measurements causes the spin to undergo a quantum dynamics phase transition from coherent trapping to coherent oscillation. Single-spin NMR at room-temperature with a spectral resolution of 3.8 Hz is achieved. These results enable the use of measurement-correlation schemes for the detection of very weakly coupled single spins.**

Quantum measurement induces back-action on the measured system, causing the system to collapse into different states determined by the random measurement outcome. To monitor the dynamics of a quantum object by sequential measurement, the back-action would induce inevitable disturbance. Weak measurements, as introduced theoretically[1–7] and demonstrated experimentally with NV centers[8,9], superconducting qubits[10–12] and other systems[13,14], are a potential solution to approach the limit of negligible disturbance of the system under study. However, this comes at the price of less information on the system. NV centers in diamond have recently been shown to be exceptional sensors for nanoscale nuclear magnetic resonance (NMR)[15–22]. In these experiments, the NV center typically probes nuclear sample spins in a Ramsey-type experiment, i.e., a free evolution of the spins in between the preparation and readout steps. High-spectral resolution, necessary for chemical identification of molecules, requires long free evolution times. To avoid measurement back-action on the sample spins, previous schemes usually avoid measurements during free evolution of the system[18,23–30]. On the other hand, long free evolution times leave the sensor idle and hence decrease sensitivity. Continuous measurements, on the other hand, make optimum use of the sensor sensitivity but generate significant quantum back action on the sample spins, making back action a limitation to spectral resolution for small number of sample spins. By interleaving sequential measurements and free evolution of the system with precise timing of the measurement sequence, back-action can be mitigated while at the same time sensitivity can be increased drastically. In addition, if the measurement rate is larger than the hyperfine coupling to nuclear spins, an effective decoupling of the nuclear spin precession from the electron spin[27,28] can be achieved, which further reduces the extra decoherence of the target nuclear spin induced by the senor decay.

Exploiting these features of sequential weak measurements, we devise theoretically and demonstrate experimentally a scheme to reconstruct the time-evolution of a single nuclear spin from the random results of many subsequent measurements by carefully tuning the strength and timing of measurements. By analyzing the correlations in the measurement record we reach the limit where measurements on single quantum systems are nearly back-action free and reconstruct their dynamics with high precision.

Weak measurements are implemented as shown in Fig. 1a. The sensor spin is initially polarized along the $x$-axis, denoted as $|x\rangle$. Essentially, the measurement of the nuclear spin consists of a rotation of the electron spin conditional on the quantum state of the nuclear spin (controlled phase gate). The controlled phase gate $e^{2i\alpha\hat{S}_z\hat{I}_X}$ is based on the magnetic dipole interaction between the sensor spin $\hat{\mathbf{S}}$ and the target spin $\hat{\mathbf{I}}$, with interaction strength α (see Methods). As shown in Fig. 1b, the controlled phase gate causes the sensor spin to rotate about the $z$-axis by an angle $\pm\alpha$ for the target spin state $|\pm X\rangle$ (i.e., polarized parallel or anti-parallel to the $X$-axis). The evolution of the sensor spin for a certain initial state of the target is $|x\rangle\otimes\left(a|+X\rangle+b|-X\rangle\right)$ → $a|+\alpha\rangle\otimes|+X\rangle+b|-\alpha\rangle\oplus|-X\rangle$ with $|\pm\alpha\rangle$ denoting the sensor state rotated away from the $x$ axis by $\pm\alpha$ . The sensor spin is then measured along the $y$ axis. The probabilities of the two outcomes, $m_k=\pm1$ are depend on the initial target state $|\pm X\rangle$. The projective measurement of the sensor therefore constitutes a measurement of the target spin **I**, with a strength depending on the value of α. In particular, a projective (strong) measurement is found for $\alpha=\pi/2$. Weak measurements can also be used for heralded initialization of the target spin along the $X$-axis upon post-selecting one of the outcomes of the sensor measurement[9].

In between two successive measurements, the target spin undergoes free precession about the *Z*-axis by an angle Φ. Thus, the outcomes of any two measurements correlate depending on the spin precession (see Methods). In the limit $\alpha \to 0$, the measurement-induced dephasing (disturbance) is negligible, and the correlation function between two measurement outcomes separated by *N* measurements, $C(N) \equiv \langle m_{k+N} m_k \rangle = \sin^2 \alpha \cos(N\Phi)$, oscillates at the precession frequency Φ (in units of radian per measurement cycle) (see Methods), without measurement induced damping. In this limit, i.e., for very weak measurements the target spin precession frequency can be determined with arbitrary spectral resolution. In practice, however, each measurement dephases the target spin (in the *X* basis) and the spin dynamics shows a damped Rabi oscillation (about the *Z*-axis). By taking this finite dephasing into account, we find the expression for the correlation function

$$C(N) = \sin^2 \alpha \left( C_+ \eta_+^N + C_- \eta_-^N \right) / 2,$$

where $C_\pm = 1 \pm \mu \cos\Phi \big/ \sqrt{\mu^2 - \sin^2 \Phi}$, $\eta_\pm = \left( \cos\Phi \pm \sqrt{\mu^2 - \sin^2 \Phi} \right) \cos^2 \frac{\alpha}{2}$ with $\mu \equiv \tan^2(\alpha/2)$ (which can be regarded as the measurement strength for $\alpha \le \pi/2$).

There is a phase transition in the quantum dynamics between coherent oscillation and coherent trapping at the boundary $\mu^2 = \sin^2 \Phi$, due to the competition between the free precession and measurement-induced dephasing (similar to damped Rabi oscillations)[2,6,7,31,32].

On one hand, when the measurement is relatively weak, i.e., $\mu^2 < \sin^2 \Phi$, the correlation function is an oscillatory function with an effective decay of $\gamma_{\text{eff}} = -\frac{1}{2}\ln(\cos\alpha)$ per measurement cycle. This is half the measurement-induced dephasing rate as the measurement dephases the spin only along the *Y*-axis, while the spin is rotating in the X-Y plane. Taking this into account, the renormalized angular frequency of the target spin precession is given by (in units of radian per measurement cycle)

$$\Phi_{\text{eff}} = \arccos\frac{\cos\Phi}{\sqrt{1-\mu^2}}.$$

Due to this, the frequency is either dragged towards 0 (oscillation slowed down) or $\pi$ (oscillation sped up) depending on the precession angle $\Phi$ being less or greater than $\pi/2$ (see Fig. 1c & d).

On the other hand, when the measurement is relatively strong, i.e., $\mu^2 > \sin^2\Phi$, the correlation function exhibits an exponential decay i.e., $C(N) \sim \exp(-N\gamma_{\text{eff}})$ if $\cos\Phi > 0$ or decay with alternating sign, $C(N) \sim (-1)^N \exp(-N\gamma_{\text{eff}})$ if $\cos\Phi < 0$, with an effective decay rate $\gamma_{\text{eff}} = \min(-\ln|\eta_+|, -\ln|\eta_-|) \approx \sin^2\Phi/[2\tan^2(\alpha/2)]$. This indicates that the spin is trapped approximately along the *X* axis. The physical picture of the trapped dynamics is illustrated in Fig. 1e & f. The trapping dynamics is similar to the quantum Zeno effect[31,32], where the trapped states are coherent superpositions of the energy eigenstates $|\pm Z\rangle$ – the eigenstates of the free precession.

In our experiments, we use the electron spin of a single NV center in diamond as the quantum probe for single $^{13}C$ nuclear spins in close proximity. Experiments are carried out at room-temperature, where the spin lifetime of the NV center spin is on the order of milliseconds. The diamond crystal in use has a $^{13}$C abundance depleted to 0.005%. The electron spin $\hat{\mathbf{S}}$ and the $^{13}$C nuclear spin $\hat{\mathbf{I}}$ are coupled via the hyperfine interaction $\hat{S}_z\mathbf{A}\cdot\hat{\mathbf{I}}$, where the $z$ axis is along the NV symmetry axis and the hyperfine interaction strength $A$ ~ kHz. We apply an external magnetic field $B \approx 2{,}561$ Gauss along the $z$ axis, and choose the transition between $|m_S = 0\rangle$ and $|m_S = -1\rangle$ as the sensor qubit.

The control scheme for one measurement cycle is shown in Fig. 1g. The electron spin is optically pumped into the state $|0\rangle \equiv |m_S = 0\rangle$ and then rotated to the $x$ axis state $|+x\rangle$ by a π/2 pulse about the $y$ axis. The $y$ component of the electron spin is measured by applying a (−π/2) pulse around the $x$ axis followed by a projective measurement along the $z$ axis via optical excitation and fluorescence detection. The correlation function is extracted from the photon count statistics (see Methods and SI). Between the initialization and readout of each cycle, we apply a sequence of $N_{\mathrm{p}}$ equally spaced π-pulses, i.e., a dynamical decoupling (DD) control on the electron spin to modulate the hyperfine interaction for a total duration of $t_{\mathrm{I}} = N_{\mathrm{p}}\tau$ (with $\tau$ being the pulse interval). The Knill DD sequence KDD*n*, which contains *n* units of 20 pulses applied along different axes (Fig. 1g) is chosen, to tolerate pulse errors in the many-pulse DD control[33].

The evolution during the DD can be factorized into a control-phase gate and a free precession of the nuclear spin $\hat{U} = \exp\left(-i\Phi\hat{I}_Z\right)\exp\left(2i\alpha\hat{S}_z\hat{I}_X\right)$. If the hyperfine interaction is weak, the precession angle $\Phi \cong 2\pi\left|\gamma_{^{13}C}\mathbf{B} - \mathbf{A}/2\right|t_{\mathrm{I}} \equiv 2\pi\bar{\nu}t_{\mathrm{I}}$ (with the $^{13}$C gyromagnetic ratio $\gamma_{^{13}\mathrm{C}} = -1{,}070.5$ Hz/Gauss and $\bar{\nu}$ denoting the hyperfine-renormalized Larmor frequency) and the conditional phase shift $\alpha \cong \frac{2A_{\perp}}{\bar{\nu}}\left|\frac{\sin\left(N_{\mathrm{p}}\pi\bar{\nu}\tau\right)}{\cos\left(\pi\bar{\nu}\tau\right)}\right|\sin^{2}\frac{\pi\bar{\nu}\tau}{2}$ (with $\mathbf{A}_{\perp}$ denoting hyperfine interaction in the *X*-*Y* plane)[9,23–25,34].

The conditional phase shift $\alpha$ and hence the measurement strength is controllable by changing the DD timing and length. In particular, if the resonant DD condition $2\tau \approx 1/\bar{\nu}$ is satisfied, $\alpha \cong 2N_{\mathrm{p}}A_{\perp}\tau = 2A_{\perp}t_{\mathrm{I}}$, which is proportional to the number of DD pulses[9,23–25,34].

The free precession angle per measurement cycle $\Phi$ is also controllable by inserting a waiting time between DD control of neighboring cycles. For resonant DD, $\Phi \approx 0 \bmod (2\pi)$. If the magnetic field is much stronger than the hyperfine interaction, i.e., $\gamma_{^{13}\mathrm{C}}B >> A$, which is the case in our experiment since the former is ~ MHz and the latter ~ kHz, the *Z* axis of the free precession is nearly the same as the *z* axis (the magnetic field direction). The free precession angle per measurement cycle then becomes $\Phi = 2\pi\bar{\nu}t_{\mathrm{c}} \bmod 2\pi$, where $t_{\mathrm{c}}$ is the duration of a measurement cycle including the DD duration $t_{\mathrm{I}}$, the readout and initialization time, and the waiting time.

Figure 2 shows the control of the measurement strength relative to the precession rate of a $^{13}$C spin weakly coupled to an NV center. The nuclear spin Larmor frequency $\nu_0 = \left|\gamma_{^{13}\mathrm{C}}B\right| \approx 2.743189$ MHz. The interval between neighboring DD pulses is set to satisfy the near-resonance condition ($2\nu_0\tau \approx 1.0001$). The oscillation frequency of the correlation function $\nu_{\rm eff}$ is determined relative to the bare Larmor frequency $\nu_0$. The experimental data is reproduced well in theory with the hyperfine interaction parameters $A_Z = 1.144$ kHz and $A_\perp = 16$ kHz. In Fig. 2a the number of DD pulses is $N_{\rm p} = 40$ . The measurement strength is chosen small compared to the free precession ($\alpha \approx 0.0743\pi$ << $\Phi \approx 0.445\pi$ ). The correlation function oscillates coherently with slow decay (Fig. 2a), and the spectral peak is close to the Larmor frequency $\bar{\nu} \approx \nu_0 +$ 0.571kHz, with a small broadening due to the measurement-induced dephasing (Fig. 2d). By increasing the number of pulses ( $N_{\rm p} = 100$ ), and hence the measurement strength ( $\alpha = 0.189\pi$ ), while keeping the free precession angle per measurement cycle $\Phi$ nearly invariant, the decay of the oscillating correlation function (Fig. 2b) and the spectral broadening (Fig. 2e) become more significant. Making the precession angle $\Phi$ close to $\pi$ by adjusting the cycle duration (the theoretical value of $\Phi$ is about $0.990\pi$), while keeping the measurement strength the same ( $\alpha = 0.189\pi$ ), we observe the correlation function to decay exponentially with alternating sign (Fig. 2c). The spectral peak is pinned at $1/(2t_{\rm c})$, about 246 Hz relative to $\bar{\nu}$ mod $1/t_c$ (see Fig. 2f), which indicates that the nuclear spin dynamics is coherently trapped.

The phase transition in quantum dynamics between coherent oscillation and coherent trapping of the nuclear spin can be seen in Fig. 3. The phase boundary $\mu^2 \equiv \tan^4(\alpha/2) = \sin^2\Phi$ is indicated by the white lines in Figs. 3a & 3b, which present the dependence of (a) the effective angular frequency $\Phi_{\rm eff} = 2\pi\nu_{\rm eff}t_{\rm c}$ and (b) the decay per measurement cycle $\gamma_{\rm eff}$ of the correlation function on the measurement strength (in terms of $\alpha$) and the nuclear spin free precession frequency (in terms of $\Phi = 2\pi\bar{\nu}t_{\mathrm{c}}$). For relatively weak measurements ($\mu^2 < \sin^2\Phi$), the nuclear spin performs coherent oscillations; otherwise, the nuclear spin is coherently trapped along one direction (for $\cos\Phi > 0$) or alternating in opposite directions (for $\cos\Phi < 0$).

In the experiment, by fixing the number of DD pulses (and hence the measurement strength $\alpha$), the phase transition is observed by varying the measurement cycle duration $t_{\mathrm{c}}$ (hence $\Phi$). Examples are shown in Figs. 3c & 3d for two different measurement strengths (corresponding to the short horizontal lines in Figs. 3a & 3b). The effective frequencies and decay rates of the correlation function agree well with the theoretical predictions (with two fitting parameters $A_Z$ and $A_\perp$ the same as in Fig. 2). The transition between frequency dragging and trapping can be seen clearly, and so is a sudden change in the derivative of the decay rate.

The measurement-induced decay (and resonance broadening) can be made arbitrary small by choosing an arbitrarily small measurement strength. In the weak measurement limit ($\alpha \to 0$ and $\mu^2 << \sin^2\Phi$), the frequency dragging is negligible, $\nu_{\mathrm{eff}} \cong \bar{\nu} + \bar{\nu}\,\mu^2/\left(2\sin^2\Phi\right)$, and the spectral resolution, limited by the measurement-induced broadening, is $\delta\nu = \gamma_{\mathrm{eff}}/\left(2\pi t_{\mathrm{c}}\right) \cong \alpha^2/\left(8\pi t_{\mathrm{c}}\right)$.

To demonstrate spectral resolution beyond the $1/T_1$ limit of the sensor electron spin, we choose an NV center (referred to as NV2) in the same diamond crystal as used for Figs. 2 & 3 but with weaker coupled $^{13}$C nuclear spins. To enhance the photon count contrast between different states of the NV center spin, we perform repetitive readout of the electron spin for 40 times assisted by the $^{14}$N nucleus[35,36] during the waiting time (see Methods and SI).

The correlation spectrum for $N_{\mathrm{p}} = 100$, as shown in Fig. 4a, exhibits a narrow peak, with half width at half maximum (HWHM) of 1.9 Hz (C1), and a broader one (C2), with HWHM of 8.75 Hz, both well below the $1/T_1$ limit (about 80 Hz) of the NV center electron spin (which has $T_1 \approx 2$ ms). The two resonances are ascribed to two $^{13}$C nuclear spins, with longitudinal hyperfine coupling $A_Z = 19 \pm 2$ Hz (C1) and $A_Z = 178 \pm 8$ Hz (C2). By increasing the number of DD pulses to $N_{\mathrm{p}} = 300$ and hence enhancing the measurement strength (Fig. 4b), the two resonances are broadened. The dependence of the HWHM, as shown in Fig. 4c, agrees well with the measurement-induced broadening plus an additional broadening $\Gamma_0$,

$$\delta\nu = \frac{\alpha^2}{8\pi t_{\mathrm{c}}} + \Gamma_0 = \frac{A_{\perp}^{2} t_I^{2}}{2\pi t_c} + \Gamma_0 .$$

The transverse hyperfine $A_{\perp}$ is fitted to be about 2.2 kHz (C1) and 4.05 kHz (C2), and the additional broadening $\Gamma_0$ is fitted to be about 0.29 Hz (C1) and 6.24 Hz (C2). In addition to intrinsic broadening (due to, e.g., dipolar interaction with other nuclear spins), one possible contribution to the additional broadening $\Gamma_0$ is the hyperfine coupling fluctuation due to the NV center electron jumping randomly among different levels during readout (details in SI). It is estimated to be $\Gamma_0^{\mathrm{hf}} \approx A_Z^2 \tau_{\mathrm{eff}}^2 / (2\pi t_{\mathrm{c}})$, with $\tau_{\mathrm{eff}}$ denoting the effective period in each cycle, during which the NV center state is in a random state ($\tau_{\mathrm{eff}}$ : 600 μs in the experiment of Fig. 4). $\Gamma_0$ of C2 is about 30 times greater than that of C1, which is consistent with the larger $A_z$ of C2. From Fig. 4c it is clear, that the spectral resolution in all the cases studied is limited by the measurement-induced dephasing.

A key aspect of the weak measurement protocol is the data acquisition time $T^{\mathrm{D}}$ for achieving a given spectral resolution $\Delta\nu$ (see Supplementary Information)

$$T^{\mathrm{D}} \propto \frac{1}{\Delta\nu}$$

For comparison, to achieve a resolution beyond the limit set by the lifetime of the sensor, a Ramsey-like scheme is also possible, in which the target spin undergoes an initialization-precession-measurement process in each readout step. The precession time between the initialization and the measurement determines the spectral resolution. However, for a weakly coupled nuclear spin with hyperfine coupling strength $A_{\perp} < 1/(2T_1)$ ($T_1$ denoting the sensor spin relaxation time), a regime of interest in this paper, the data acquisition time in the Ramsey scheme is larger than that of the sequential weak measurement scheme by a factor of : $1/(2A_{\perp}T_1)^4$ (see Supplementary Information) since no measurements are performed during the free precession time. For example, to resolving a $^{13}C$ nuclear spin located 6 nm away from the NV center (which has hyperfine interaction about 90 Hz), the data acquisition time of the Ramsey protocol is longer by a factor of about 1000.

The sequential weak measurement can in principle realize arbitrary spectral resolution of single spin NMR by further reducing the measurement strength (via, e.g., choice of DD sequences), and by suppressing the background broadening (via, e.g., the use of more purified crystals and target spins located further away from the sensor – see

SI for discussions on the spatial range of detection). This method will be particularly useful for NMR of single molecules on diamond surfaces where the hyperfine interaction with the sensor is very weak (e.g., 10 Hz to 0.1 kHz). Since the sensor itself does not limit the resolution, it is also applicable to other solid state spin systems. The sensing efficiency can be enhanced further by employing fast and efficient methods for electron readout[37]. The coherent oscillation and trapping dynamics, tunable by DD control of the central electron spins, can be exploited to control and initialize remote nuclear spins, respectively. Now that schemes are available to spectrally resolve, initialize, and coherently control multiple nuclear spins that are not required to be located closely to a central electron spin, quantum information processing with a relatively large number (e.g., >10) of nuclear spins is a step closer[38].

## METHODS

### Weak measurement formalism and correlation function

The mutually controlled phase gate is $e^{2i\alpha\hat{S}_z\hat{I}_X}$, where the coordinate axes $(x, y, z)$ of the sensor and those of the target $(X, Y, Z)$ are not necessarily identical. The sensor spin is initially in the state $|x\rangle$ and at the end measured along the *y* axis. The weak measurement is characterized by the Kraus operators $\hat{M}_{\pm} = \langle \pm y | e^{i2\alpha\hat{S}_z\hat{I}_X} | x\rangle = \left(e^{i\alpha\hat{I}_X} \pm ie^{-i\alpha\hat{I}_X}\right)/2$. Given the initial state of the target spin as described by a density matrix $\hat{\rho}$, the probability of the output $m_k = \pm 1$ is $p_{\pm} = \mathrm{Tr}\left(\hat{M}_{\pm}\hat{\rho}\hat{M}_{\pm}^{\dagger}\right) = \frac{1}{2} \pm I_X \sin\alpha$ with $I_X = \mathrm{Tr}\left(\hat{\rho}\hat{I}_X\right)$ and the state after the measurement is $\hat{\rho}_{\pm} = \left(\hat{M}_{\pm}\hat{\rho}\hat{M}_{\pm}^{\dagger}\right)/p_{\pm}$. In particular for a fully unpolarized initial state $\hat{\rho} = 1/2$, the post-measurement state is $\hat{\rho}_{\pm} = \frac{1}{2} \pm \hat{I}_X \sin\alpha$ corresponding to the output $m_k = \pm 1$, with partial polarization $\pm\sin\alpha$ along the *X* axis, which is called heralded initialization. If the output is discarded, the target state becomes

$$\hat{\mathcal{M}}[\hat{\rho}] \equiv \hat{M}_{+}\hat{\rho}\hat{M}_{+}^{\dagger} + \hat{M}_{-}\hat{\rho}\hat{M}_{-}^{\dagger} = \hat{\rho}\cos^2\frac{\alpha}{2} + 4\hat{I}_X\hat{\rho}\hat{I}_X\sin^2\frac{\alpha}{2},$$

which reduces the spin polarization in the *Y*-*Z* plane by a factor of $\cos\alpha$, that is, pure dephasing in the *X* basis (which reduces the spin polarization along the *Y* and *Z* axes, but keeps the *X* component unchanged).

After the measurement, the target spin can undergo a free precession, e.g., about the *Z* axis via the evolution $\hat{\mathcal{U}}[\hat{\rho}] = e^{-i\Phi\hat{I}_Z}\hat{\rho}e^{i\Phi\hat{I}_Z}$. The correlation function is

$$C(N) \equiv \langle m_{k+N} m_k \rangle = \mathrm{Tr}\left[ \hat{\mathcal{P}} \left( \hat{\mathcal{V}} \hat{\mathcal{M}} \right)^{N-1} \hat{\mathcal{V}} \hat{\mathcal{P}} [\hat{\rho}] \right],$$

where the polarization operator $\hat{\mathcal{P}}[\hat{\rho}] \equiv \hat{M}_+ \hat{\rho} \hat{M}_+^\dagger - \hat{M}_- \hat{\rho} \hat{M}_-^\dagger$ denotes the heralded initialization.

The precession of the target spin polarization **I** is described by the transform

$$\mathcal{V} \begin{pmatrix} I_X \\ I_Y \\ I_Z \end{pmatrix} = \begin{pmatrix} \cos\Phi & -\sin\Phi & 0 \\ \sin\Phi & \cos\Phi & 0 \\ 0 & 0 & 1 \end{pmatrix} \begin{pmatrix} I_X \\ I_Y \\ I_Z \end{pmatrix}.$$

The measurement transforms the polarization by

$$\mathcal{M} \begin{pmatrix} I_X \\ I_Y \\ I_Z \end{pmatrix} = \begin{pmatrix} 1 & 0 & 0 \\ 0 & \cos\alpha & 0 \\ 0 & 0 & \cos\alpha \end{pmatrix} \begin{pmatrix} I_X \\ I_Y \\ I_Z \end{pmatrix}.$$

The eigenvalues of the transformation $\mathcal{V}\mathcal{M}$ are easily obtained as $\eta_Z = \cos\alpha$ and $\eta_\pm = \left( \cos\Phi \pm \sqrt{\mu^2 - \sin^2\Phi} \right) \cos^2 \frac{\alpha}{2}$ with $\mu \equiv \tan^2(\alpha/2)$, corresponding to the right eigenvectors $\mathbf{v}_Z^{\mathrm{R}} = (0,0,1)^{\mathrm{T}}$ and $\mathbf{v}_\pm^{\mathrm{R}} = \left( \cos\Phi \sin^2 \frac{\alpha}{2} \pm \Delta, \sin\Phi, 0 \right)^{\mathrm{T}}$ with $\Delta = \cos^2 \frac{\alpha}{2} \sqrt{\mu^2 - \sin^2\Phi}$, and left eigenvectors $\mathbf{v}_Z^{\mathrm{L}} = (0,0,1)$ and $\mathbf{v}_\pm^{\mathrm{L}} = \frac{1}{2\Delta \sin\Phi} \left( \pm\sin\Phi, \mp\cos\Phi \sin^2 \frac{\alpha}{2} - \Delta, 0 \right)$. They satisfy the orthonormal conditions $\mathbf{v}_i^L \mathbf{v}_j^R = \delta_{ij}$ and $\mathbf{v}_+^{\mathrm{R}} \mathbf{v}_+^{\mathrm{L}} + \mathbf{v}_-^{\mathrm{R}} \mathbf{v}_-^{\mathrm{L}} + \mathbf{v}_Z^{\mathrm{R}} \mathbf{v}_Z^{\mathrm{L}} = 1$. For the target spin initially in the fully unpolarised state, the heralded initialization polarizes the spin along the *X*-axis to be $(\sin\alpha, 0, 0)^{\mathrm{T}} = \sin\alpha \mathbf{e}_X$ and the following measurement and precession keeps it in the *X*-*Y* plane. So only the last two eigenstates of $\mathcal{V}\mathcal{M}$ are relevant. The correlation function is

$$C(N)=\sin^2\alpha\left[\left(\mathbf{e}_X^{\mathrm{T}}\mathbf{v}_+^{\mathrm{R}}\mathbf{v}_+^{\mathrm{L}}\mathbf{e}_X\right)\eta_+^N+\left(\mathbf{e}_X^{\mathrm{T}}\mathbf{v}_-^{\mathrm{R}}\mathbf{v}_-^{\mathrm{L}}\mathbf{e}_X\right)\eta_-^N\right]$$
$$=\sin^2\alpha\left(\frac{\eta_+^N+\eta_-^N}{2}+\frac{\eta_+^N-\eta_-^N}{\eta_+-\eta_-}\cos\Phi\sin^2\frac{\alpha}{2}\right).$$

1) If $\mu^2<\sin^2\Phi$, the correlation function oscillates with a dragged frequency and an effective decay;
2) If $\mu^2>\sin^2\Phi$ and $\cos\Phi>0$, the correlation function for large *N* is dominated by $\eta_+^N$, an exponential decay corresponding to trapping along the axis rotated from the *X* axis by an angle $\approx\Phi$ about the *Z* axis;
3) For $\mu^2>\sin^2\Phi$ and $\cos\Phi<0$, the correlation function for large *N* is dominated by $\eta_-^N$, an exponential decay with alternating sign $(-1)^N$ corresponding to coherent trapping alternatively parallel and anti-parallel to the axis rotated from the *X* axis by an angle $\approx\Phi$ about the *Z* axis.

**Diamond crystal**

A 99.995% $^{12}$C-enriched diamond crystal (5.3 mm × 4.7 mm × 2.6 mm) was grown by the temperature gradient method under HPHT conditions of 5.5 GPa and 1350°C using high-purity Fe–Co-Ti solvent and high-purity $^{12}$C-enriched solid carbon. The crystal was irradiated by 2 MeV electrons at room temperature to the total fluence of 1.3 x $10^{11}$ cm$^{-2}$ and annealed at 1000°C (for 2 hours in vacuum) to create single NV centers from intrinsic nitrogen impurities. A polished, (111)-oriented slice (2 mm × 2 mm × 80 μm) obtained by laser-cutting has been used in the present work. The isotopic enrichment enables the detection of single, weakly coupled $^{13}$C nuclear spins, and mitigates the existence of a strong, overlapping $^{13}$C spin bath. The $T_2^*$ is typically on the order of 50 μs.

**Experimental setup**

The diamond crystal is positioned inside a superconducting vector magnet ($B_z = 3$ T, $B_{x,y} = 0.2$ T), with the diamond surface normal pointing along the main magnetic field axis. The magnet is running at a field of about $B$ = 2,561 Gauss with a magnetic field stability of ~$2.5 \cdot 10^{-5}$G/h (corresponding to $^{13}$C Larmor frequency stability ~ 0.0274 Hz/h). For a typical accumulation time of ~3 h for one measurement, this results in a negligible drift of the $^{13}$C Larmor frequency of about 0.08 Hz. The $^{13}$C Larmor frequency $\nu_0 = \left|\gamma_{^{13}\mathrm{C}} B\right|$ is measured to be $2{,}743{,}189 \pm 190$ Hz for Figs. 2 & 3 and $2{,}740{,}090.4 \pm 7.9$ Hz for Fig. 4 (see SI). The magnetic field shifts the NV centre transition $\left|m_S = 0\right\rangle \leftrightarrow \left|m_S = -1\right\rangle$ to around 4.3 GHz.

The experiment consists of a home-built confocal microscope with a 520 nm excitation laser diode. The laser can be switched on and off on the timescale of 10 ns. The photoluminescence of single NV centres is collected via an oil-immersion objective with a numeric aperture of 1.35 and detected by an avalanche photo-diode, capable of detecting single photons. The spin resonance is detected optically via spin state dependent fluorescence of single NV centres. Microwave (MW) radiation is generated by an arbitrary waveform generator (AWG) with a sampling rate of 12 GS/sec, and subsequent amplification up to a power of around 40 dBm. The same AWG also controls the timing of the experiment. The microwaves are guided through coaxial cables and a coplanar waveguide, with a width of around 100 μm at the position of the NV. The RF signal used to manipulate the nitrogen nuclear spin is directed through the same waveguide.

**Nuclear-spin assisted readout**

One single readout of the NV centre spin, via a 300 ns laser pulse, produces much less than one photon on average. The readout efficiency can be increased, by transferring the spin state from the NV centre electron spin to the NV centre $^{14}N$ nuclear spin, which can subsequently be read out multiple times[35,36]. Since only the $m_S = 0$ and $m_S = -1$ spin levels of the NV centre spin are used for sensing, the $^{14}N$ nuclear spin consequently has to be initialized into a two-level manifold (in this work $m_I = 0$ and $m_I = +1$) before every measurement.

**Construction of correlation function from photon statistics**

Photon counts in experiments are not a perfect measurement of the NV center spin. We use $D(n_k \mid m_k)$ to denote the probability of detection of $n_k$ photons for the NV center spin state that would yield an output $m_k$ in a perfect measurement. The joint probability of detection of $n_k$ and $n_{k+N}$ for the $k$-th and ($k$+$N$)-th measurements is

$$p(n_k, n_{k+N}) = \sum_{m_k, m_{k+N}} D(n_k \mid m_k) D(n_{k+N} \mid m_{k+N}) \, p(m_k, m_{k+N}),$$

where the joint probability $p(m_k, m_{k+N}) = \left[1 + m_k m_{k+N} C(N)\right]/4$ depends on the correlation function. Direct calculation yields

$$C(N) = 4\left(\langle n_k n_{k+N} \rangle - \bar{n}^2\right)/(\bar{n}_+ - \bar{n}_-)^2 ,$$

where $\bar{n}_\pm$ is the averaged photon count for the senor measurement output $\pm 1$ and $\bar{n} \equiv (\bar{n}_+ + \bar{n}_-)/2$ is the average photon count.

**Acknowledgements**

We thank S. Zaiser & N. Aslam for fruitful discussions. We acknowledge financial support by the German Science Foundation (SPP1601, FOR), the EU (DIADEMS, SMel), the Max Planck Society, the Volkswagen Stiftung, the Hong Kong Research Grants Council General Research Fund (No. 14319016), the MOST of China (Grants No. 2014CB848700) and the Japan Science and Technology Agency and Japan Society for the Promotion of Science KAKENHI (Nos. 26246001 and 26220903).

**Author Contributions**

R.B.L. proposed and supervised the theoretical study; R.B.L., W.Y., P.W. & X.Y.P. formulated the theory; P.W. & W.Y. carried out the calculations; M.P., P.N. & J.W. conceived the experiments; M.P. carried out the measurements; H.S., S.O. & J.I. designed and conducted the synthesis and fabrication of the diamond substrate; M.P.,

P.W., P.N., W.Y., R.B.L. & J.W. analyzed the data; P.W., M.P., D.B.R.D., W.Y., R.B.L. & J.W. wrote the manuscript with contributions from other authors.

## Competing financial interests

The authors declare no competing financial interests.

## Data availability

Data supporting the findings of this study are available within the article and its supplementary materials and from the corresponding authors upon reasonable request.

## Code availability

Custom computer code used in the theoretical studies and experimental analysis are available from the corresponding authors upon reasonable request.

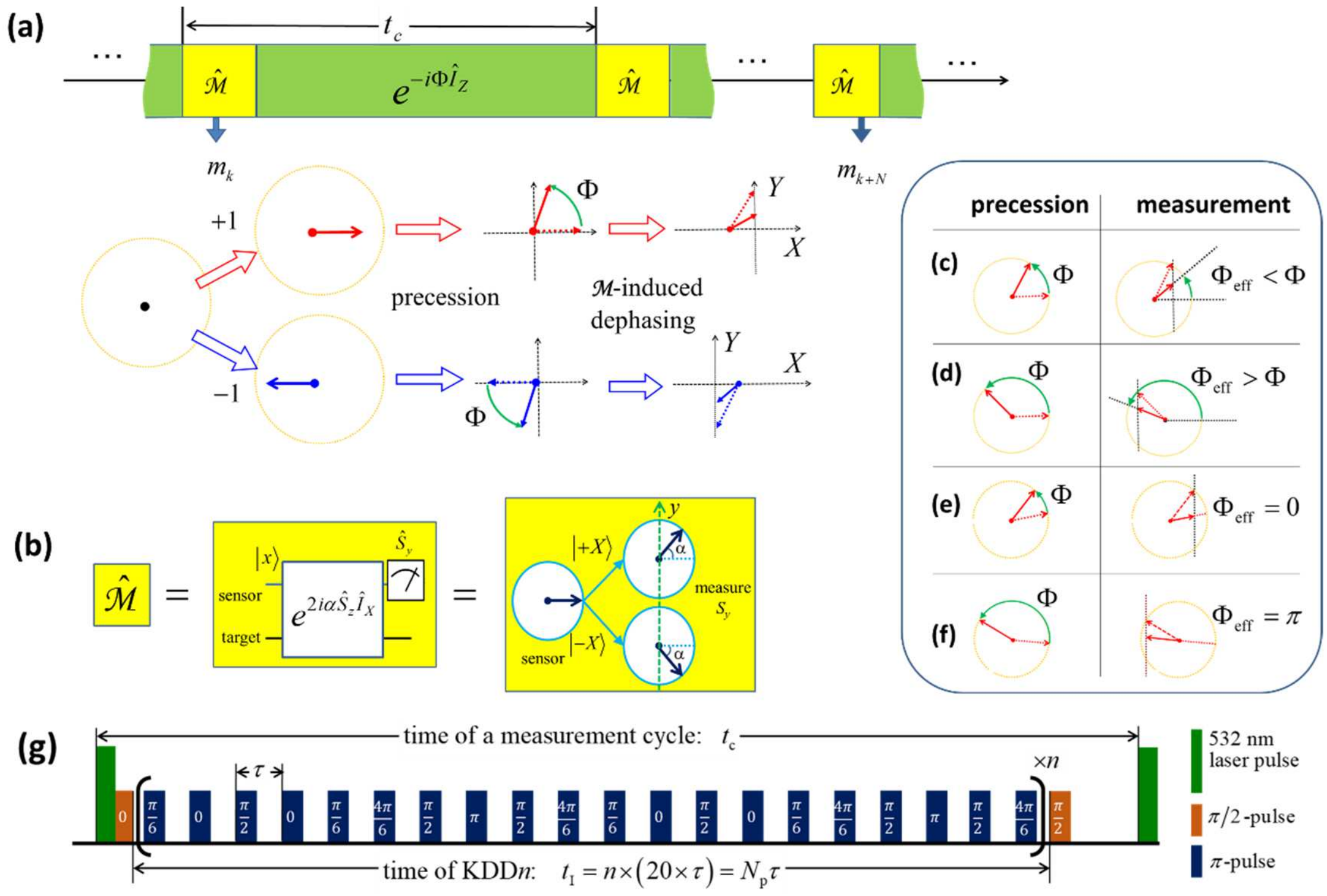

**Figure 1 | Sequential weak measurement of a target spin via projective measurement of a sensor spin.** (a) The target spin is partially polarized along $X$ or $-X$ depending on the *k*-th measurement result of the sensor spin, then undergoes free precession about the *Z* axis. If the measurement result is discarded, the target spin experiences measurement-induced dephasing along the *X* axis. The correlation between different measurements on the sensor, e.g., the *k*-th and the (*k*+*N*)-th ones, reflects the spin dynamics. (b) The sensor-assisted weak measurement ($\hat{\mathcal{M}}$) of the target spin. The sensor is initialized along the *x* axis. By the control-phase gate, the sensor spin precesses about the *z* axis by opposite angles for opposite target states $|\pm X\rangle$. A projective measurement on the sensor along the *y* axis constitutes a weak measurement of the target spin. (c-f) Precession by an angle $\Phi$ (left column) and the measurement-induced dephasing (right column), shown as evolution from dotted arrows to solid ones. (c/d) The effective precession $\Phi_{\rm eff}$ is slowed down/sped up for $\Phi$ greater/less than $\pi/2$. (e/f) For $\Phi$

close to 0 or π, the spin is coherently trapped along an axis ($\Phi_{\mathrm{eff}} = 0$) or alternatively along opposite directions ($\Phi_{\mathrm{eff}} = \pi$). (g) In the experiment, a measurement cycle contains a 532 nm laser pulse (300 ns), a KDD*n* sequence (*n* units of 20 equally separated *π*-pulses) sandwiched between two *π*/2-pulses, and a waiting time. The numbers associated with the microwave pulses indicate the angle between the rotation axes and the *y* axis.

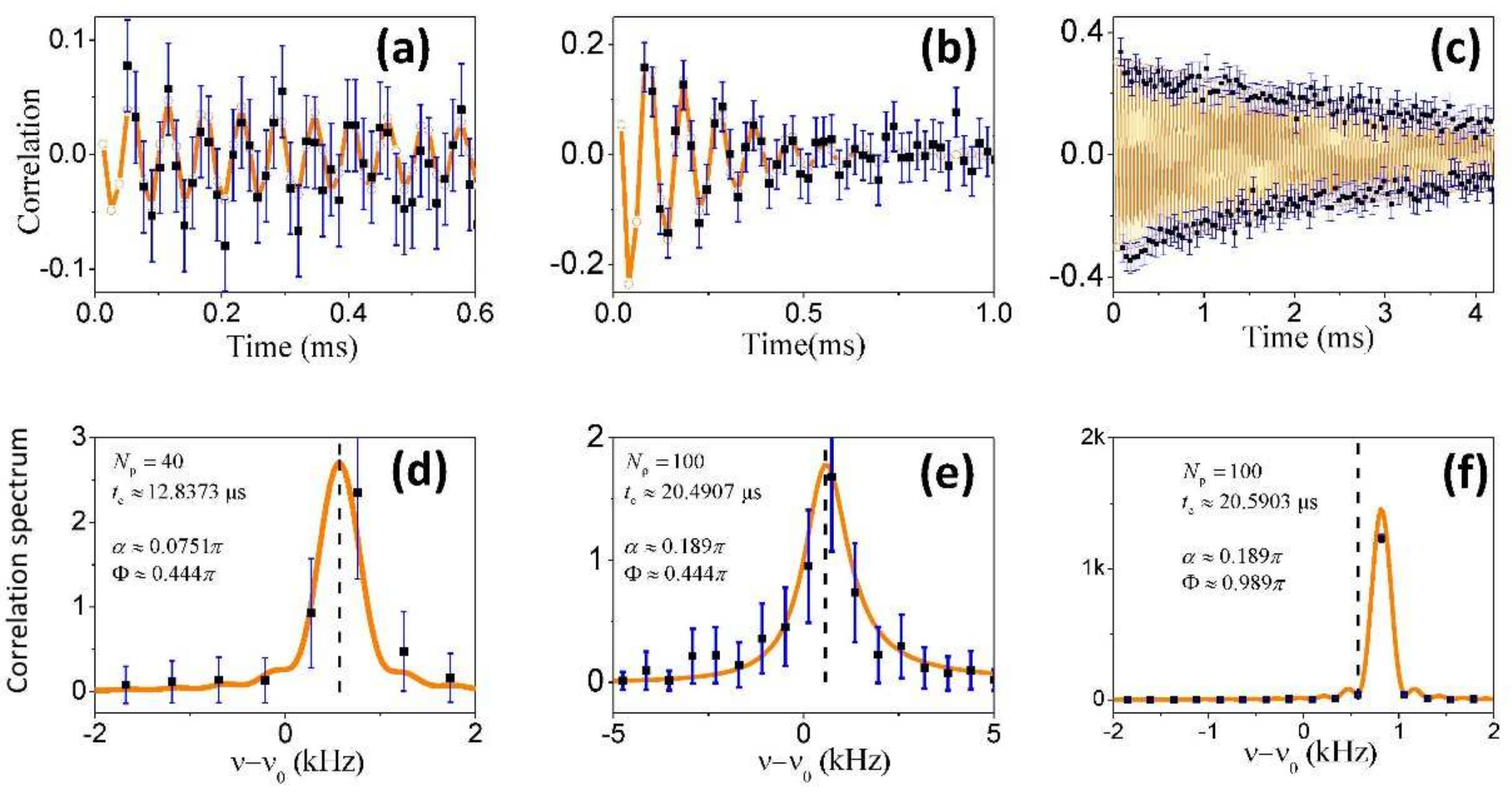

**Figure 2 | Control of the measurement strength relative to the precession frequency of a nuclear spin.** (a-c) are the correlation functions of the sequential weak measurements on the nuclear spin as functions of the delay time ($Nt_c$) between two measurements separated by $N$ cycles, with $t_c$ being the duration of each measurement cycle. The dark blue error bars represent two standard deviations ($2\sigma$) (d-f) are the corresponding correlation spectra. The experimental data are shown in black squares, and theoretical results are in orange lines and open circles. The interval between DD pulses, $\tau = 0.18224$ μs, is set near the resonance of the Larmor frequency $\nu_0 = 2.743189$ MHz. In (a/d), (b/e), and (c/f), correspondingly, the number of DD pulses is $N_p = 40$, 100, and 100, the duration of one measurement cycle is $t_c =$ 12.8373, 20.4907, and 20.5903 μs. The theoretical results reproduce the experimental data well, with the hyperfine coupling parameters $A_Z = 1.142$ kHz and $A_\perp = 16$ kHz. This yields the precession angle $\Phi =$ 0.445π, 0.443π , and 0.990π, as well as the conditional phase shift $\alpha =$ 0.0743π, 0.186π, and 0.186π, correspondingly in (a/d), (b/e), and (c/f). The vertical dashed lines in (d), (e) and (f) mark the positions of the hyperfine-modified precession frequency $\bar{\nu} - \nu_0$.

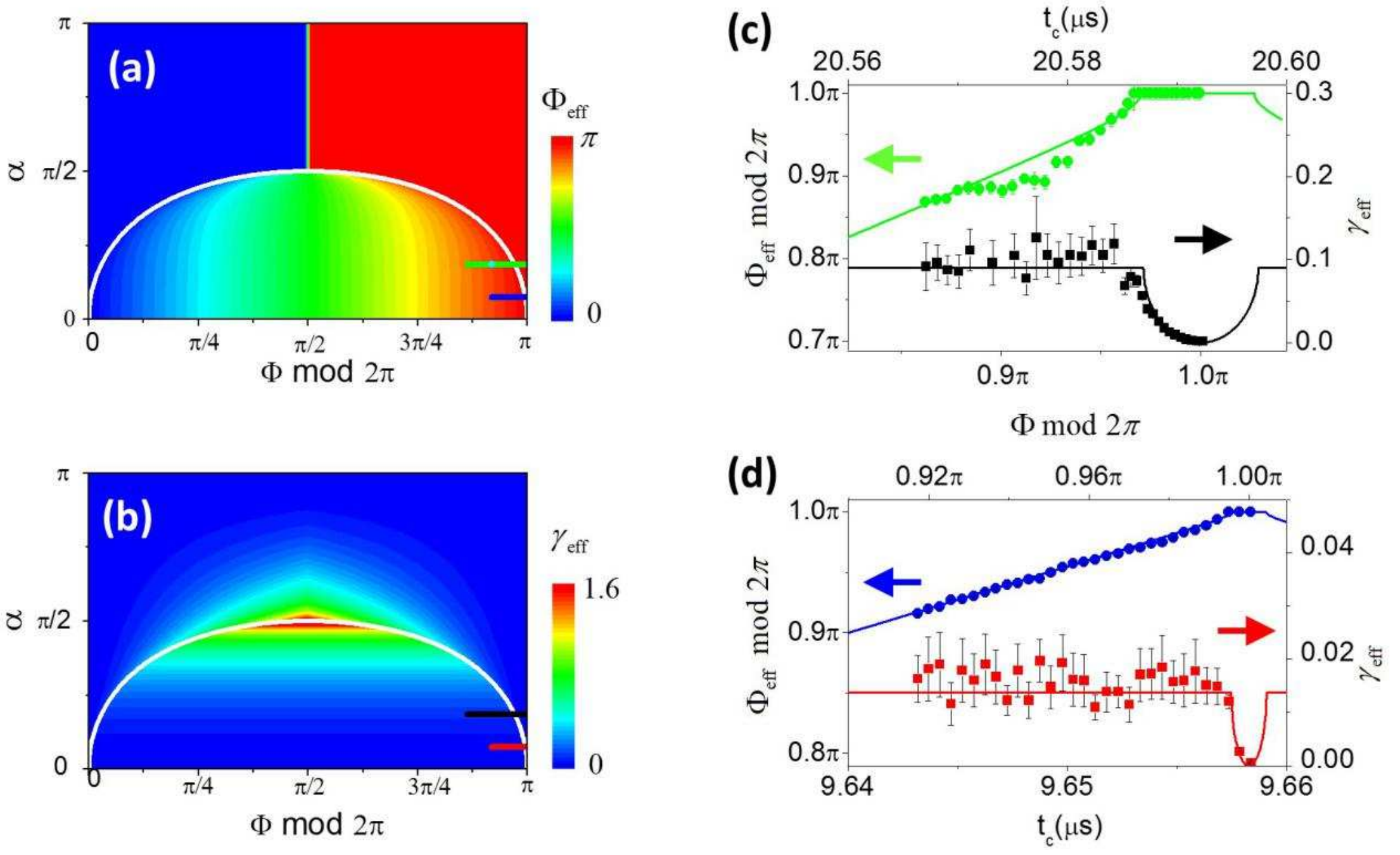

**Figure 3 | Quantum dynamics phase transition between oscillating and trapped dynamics of a nuclear spin under sequential weak measurement.** The theoretical results of (a) the effective precession angle per cycle $\Phi_{\rm eff} = 2\pi\nu_{\rm eff}t_{\rm c}$ and (b) the effective decay per cycle $\gamma_{\rm eff}$ of the correlation functions versus the conditional phase shift $\alpha$ and nuclear precession angle per measurement cycle $\Phi = 2\pi\bar{\nu}t_{\rm c}$. The white curves show the phase boundary $\tan^4(\alpha/2) = \sin^2\Phi$. The short horizontal lines indicate the parameter ranges of the experimental data in (c) and (d). (c) The experimental oscillation frequency (circles) and decay per cycle (squares) of the correlation function as functions of the measurement cycle duration $t_{\rm c}$ for KDD5 control ($N_{\rm p} = 100$), compared with the theoretical results (curves) as functions of the nuclear spin precession angle per cycle $\Phi = 2\pi\bar{\nu}t_{\rm c}$ for $\alpha$ = 0.186π. (d) The same as (c) but for KDD2 control in the experiment ($N_{\rm p} = 40$, corresponding to $\alpha$ = 0.0743π in theory). The nuclear spin, the DD sequences, and the fitting parameters $A_Z = 1.142$ kHz and $A_\perp = 16$ kHz for determining $\Phi$ and $\alpha$ in theory are the same as in Fig. 2.

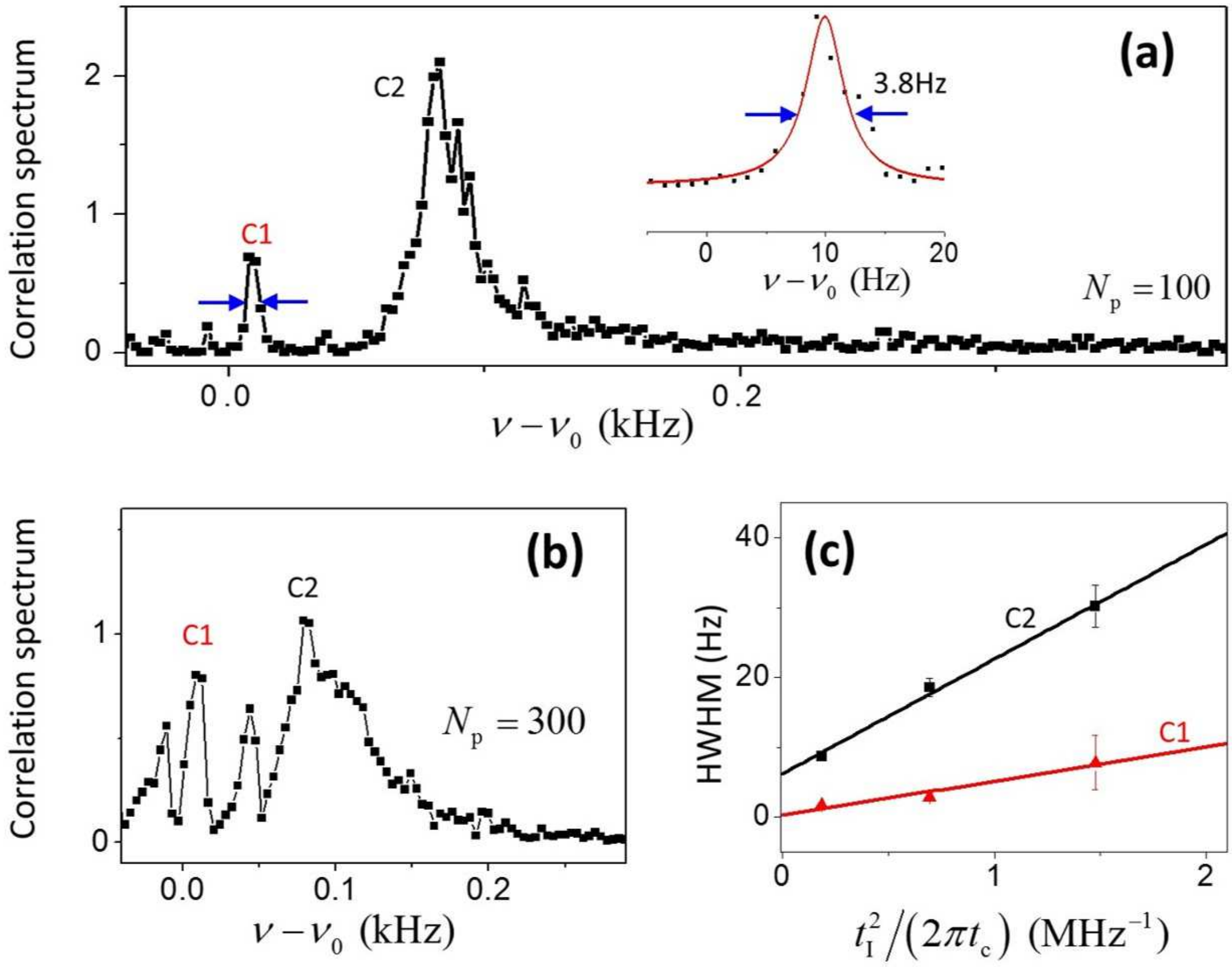

**Figure 4 | High-resolution spectroscopy of single nuclear spins.** (a) & (b) Correlation spectra of sequential measurements of single nuclear spins weakly coupled to the sensor NV2, for the KDD pulse number $N_{\mathrm{p}} = 100$ & 300, respectively. The two peaks C1 and C2 are ascribed to two different nuclear spins. The inset in (a) is a close-up of the peak C1. (c) The HWHM of the C1 and C2 resonances as functions of the interaction time squared (which is proportional to the measurement strength). The symbols are experimental data for $N_{\mathrm{p}} = 100$, 200, and 300, and the curves represent the fitted theory. The linear dependence indicates the broadening is caused mainly by measurement-induced dephasing.

# Supplementary Information

# For

"High resolution spectroscopy of single nuclear spins via sequential weak measurement"

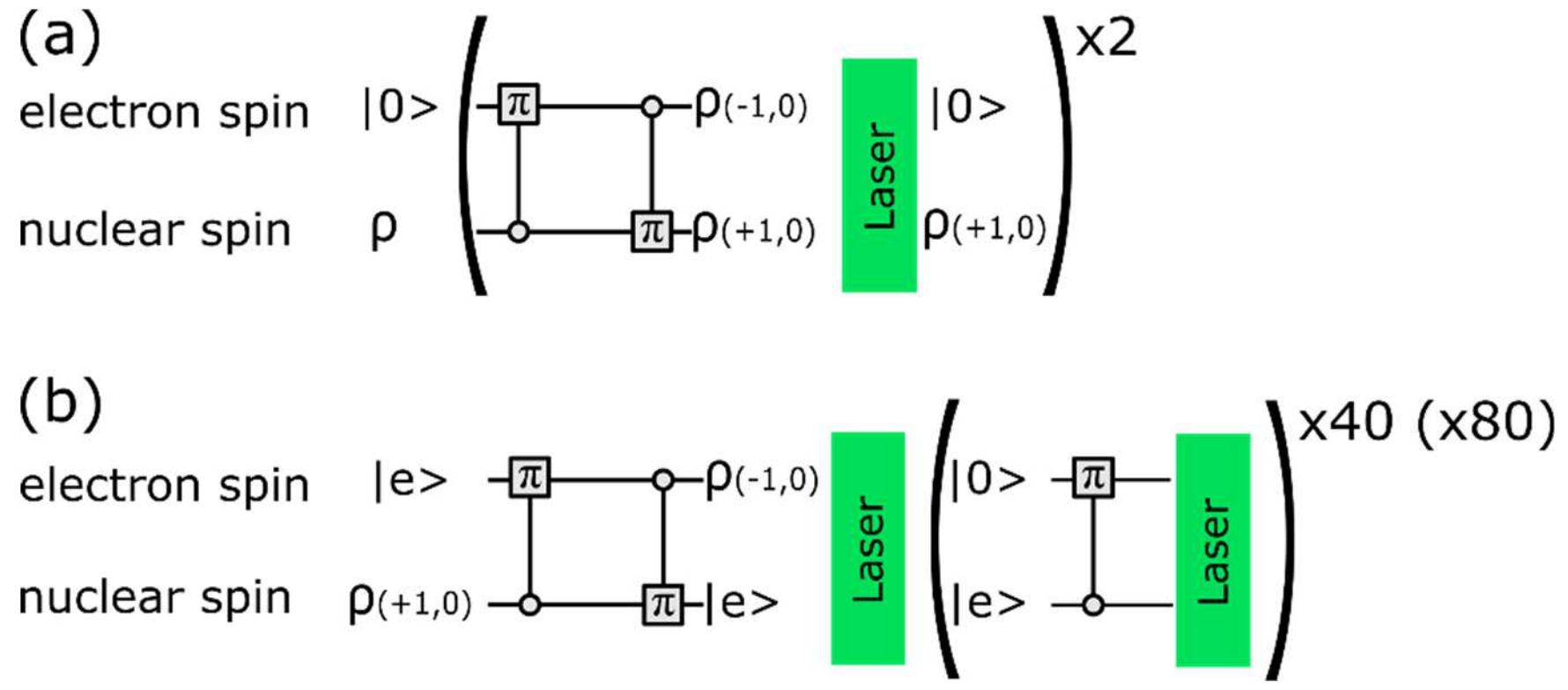

**Figure S1: Pulse sequences for repetitive readout.** (a) shows the initialization of the $^{14}N$ nuclear spin. Starting with an initialized electron and unpolarized nuclear spin, a SWAP gate consisting of two consecutive CNOT gates on the electron and nuclear spin, respectively, is performed. This initializes the $^{14}N$ nuclear spin in the desired $m_I = 0$ and $m_I = +1$ submanifold. A laser pulse then reinitializes the electron spin, leaving the nuclear spin state untouched. The whole sequence is performed twice. (b) After the measurement on the target spin, the electron spin state is read out via repetitive readout of the $^{14}N$ nuclear spin. An arbitrary electron state $|e>$ is transferred onto the $m_I = 0$ and $m_I = +1$ nuclear spin manifold by a SWAP gate. Afterwards, the nuclear spin state can be transferred onto the electron spin, and read out. Due to the stability of the $^{14}N$ state under optical illumination, this can be done repetitively (40 or 80 times in this work), effectively increasing the number of photons collected for one measurement on the target spin.

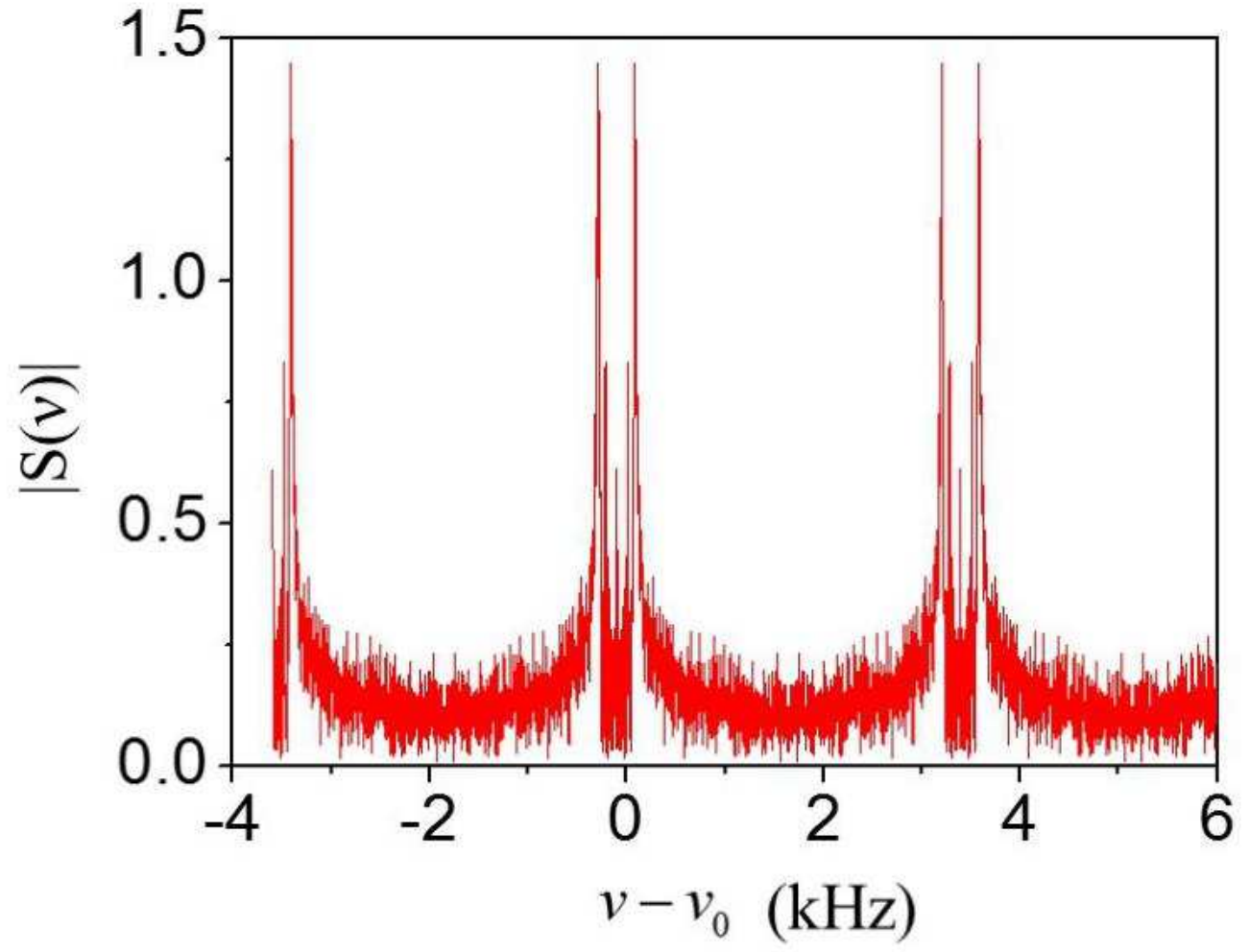

**Figure S2: Photon shot noise fluctuation of the correlation spectrum.** The fluctuation due to photon shot noises is about 0.2. The experimental condition and parameters are the same as in Fig.4(a) in the main text.

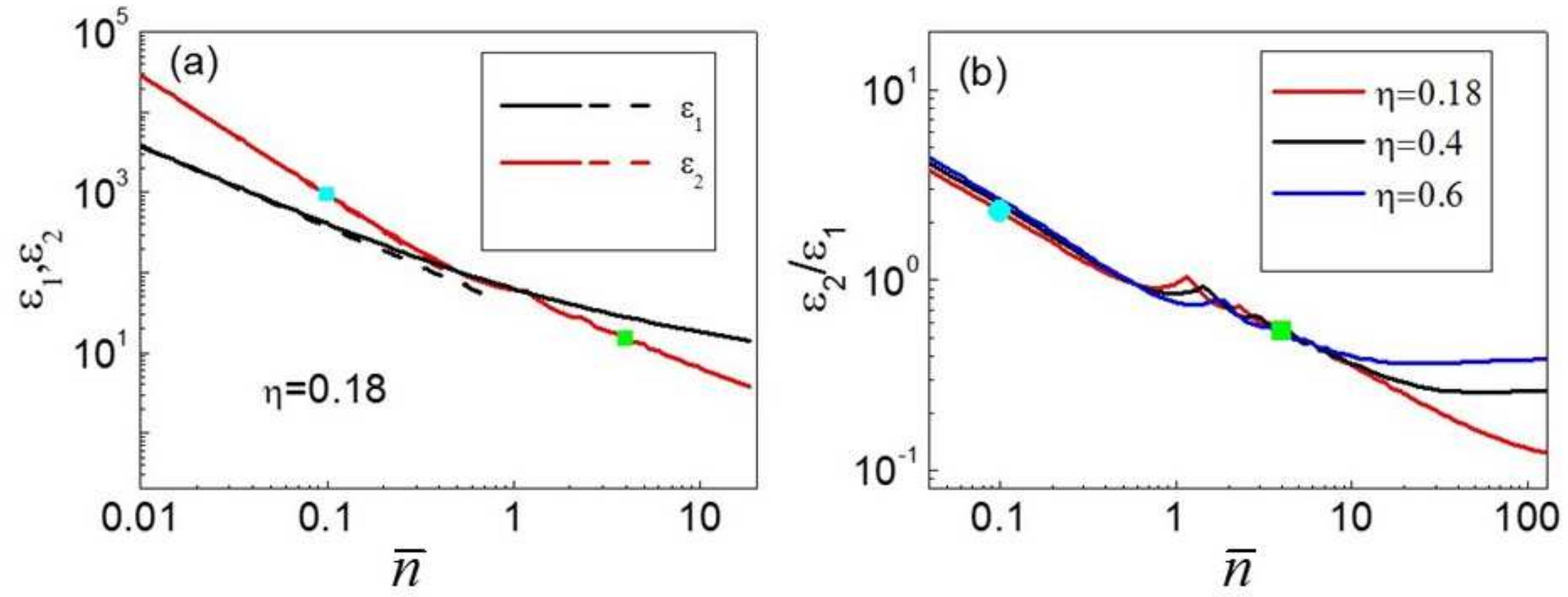

**Figure S3: Comparison between the methods for constructing correlation of weak measurements from photon counts.** (a) The fluctuation of correlations that are reconstructed with the two different methods, plot as a function of $\overline{n}$ . The photon count contrast is fixed to be 0.18. The solid line denotes the exact result while the dashed line denotes the approximate result at the limit of the low photon number. (b) The ratio between the fluctuations of the correlations obtained by the two methods, plot as a function of $\overline{n}$ . The curves of different colors correspond to different photon count contrasts $\eta$ as indicated in the legend. The parameters corresponding to Figs. 2 & 3 in the main text are marked in the figures by the Cyan symbols while those corresponding to Fig.4 are marked by the Green symbols.

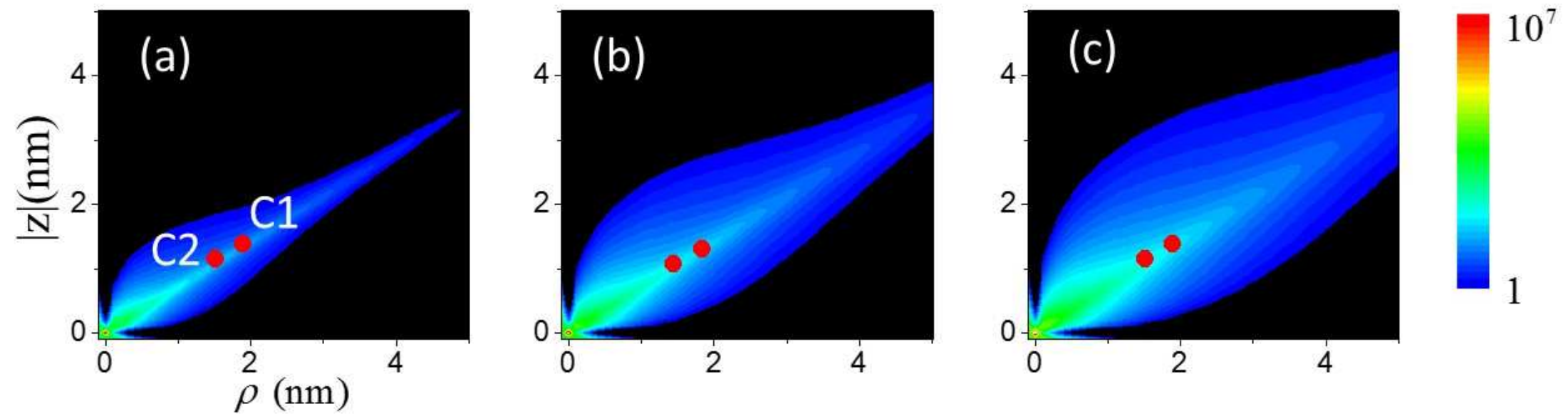

**Figure S4: Spatial range of sensing single nuclear spins**. The color contour is the signal-to-noise ratio of the resonance peak of the correlation spectrum of a $^{13}$C nuclear spin at different locations, given by the cylindrical coordinates (radius $\rho$ from the NV axis and the $z$ coordinate along the NV axis). The interaction time $t_{\mathrm{I}}$ between the NV center electron spin and the target nuclear spin is determined by the number of DD control pulses $N_{\mathrm{p}}$. The effective duration for extra dephasing is $\tau_{\mathrm{eff}} \approx 600\ \mu\mathrm{s}$. The other parameters are the same as in Figs. 4 in the main text. (a), (b), and (c): $N_{\mathrm{p}} = 100$, 200 and 300 in turn. The red spots are the locations of the nuclear spins C1 and C2 observed in Fig. 4 of the main text. For all the figures, the number of measurement cycles is $M = 3\times10^7$.

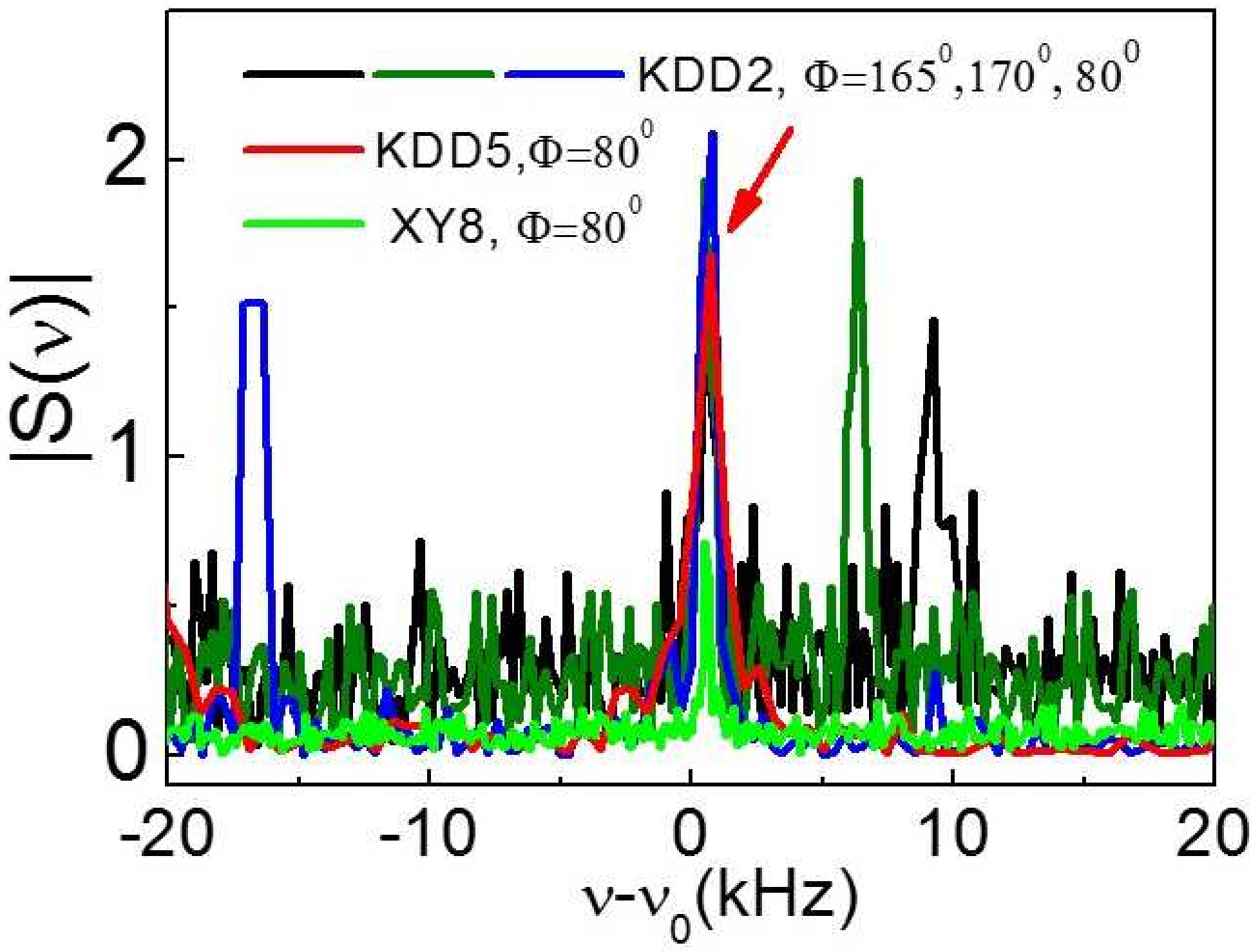

**Figure S5: Determination of the hyperfine coupling.** Correlation spectrum $|S(\nu)|$ for various interaction time $t_I$ and cycle period $t_c$. For XY8, KDD2 and KDD5, $t_I =$ 1.45920 μs, 7.29600 μs, and 18.24000 μs in turn. The Larmor frequency is $\nu_0 \approx 2.743189$ MHz (the same as in Figs. 2-4 of the main text).

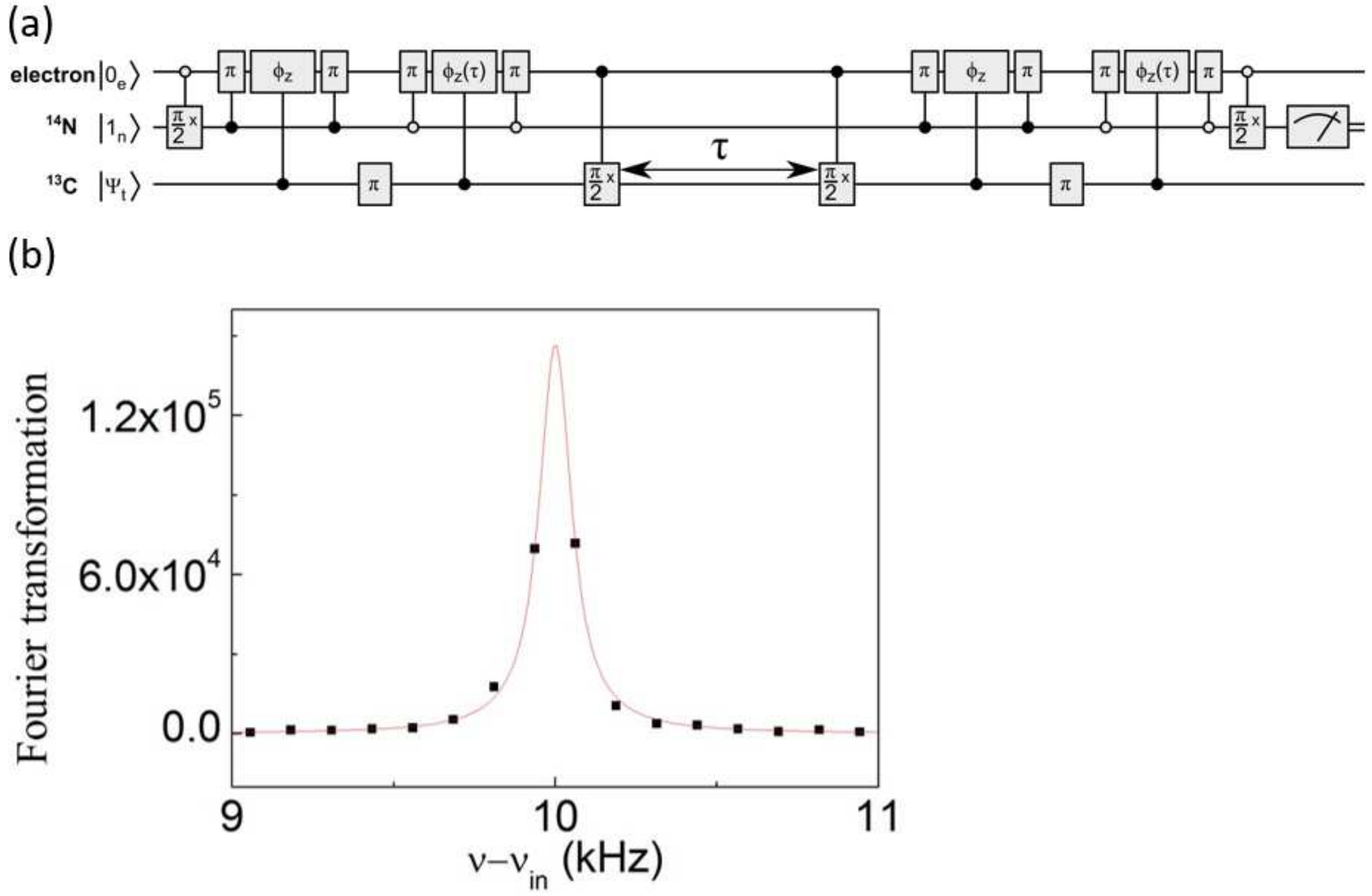

**Figure S6: High-precision measurement of the bare Larmor frequency of $^{13}$C nuclear spins.** (a) Measurement sequence to determine the bare $^{13}$C Larmor frequency by employing a hybrid spin sensor (see Ref. [1]). The sequence consists of four phase accumulation parts (denoted by $\Phi_z(\tau)$), separated by storage of the accumulated phase on the nitrogen nuclear spin, as well as radio-frequency manipulation of the $^{13}$C spins. During the manipulation, the NV centre electron spin is in the $m_S = 0$ state (hence no hyperfine interaction on the nuclear spins). The measurement is performed by varying the time $\tau$ between the two $\frac{\pi}{2}$ pulses on the $^{13}$C spins, performing an FID measurement. (b) Fourier transformation of the FID signal for measuring the bare Larmor frequency with the method in Ref. [1].

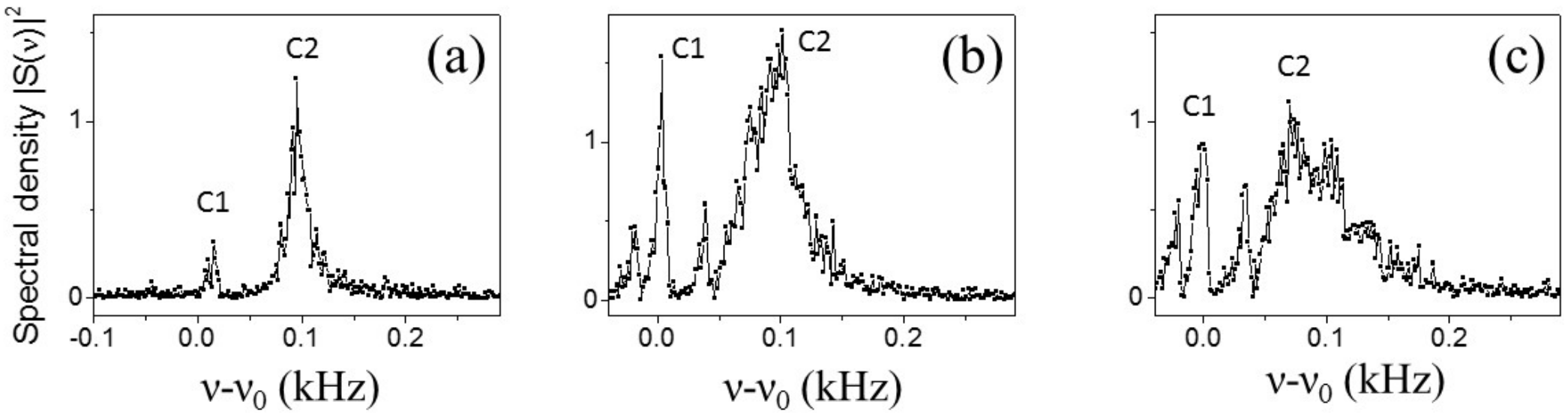

**Figure S7 | High-resolution spectroscopy of single nuclear spins (Data1).** Fourier transform of the correlation constructed from Data1. The electron spin is repetitively read out for 40 times in each measurement cycle. The number of DD pulses is (a) $N_\mathrm{p} = 100$, (b) $N_\mathrm{p} = 200$, and (c) $N_\mathrm{p} = 300$. The bare Larmor frequency is measured to be $\nu_0 = 2.740134 \pm 0.39$ Hz. The number of time points for Fourier transform is $N_{FT} = 2000$.

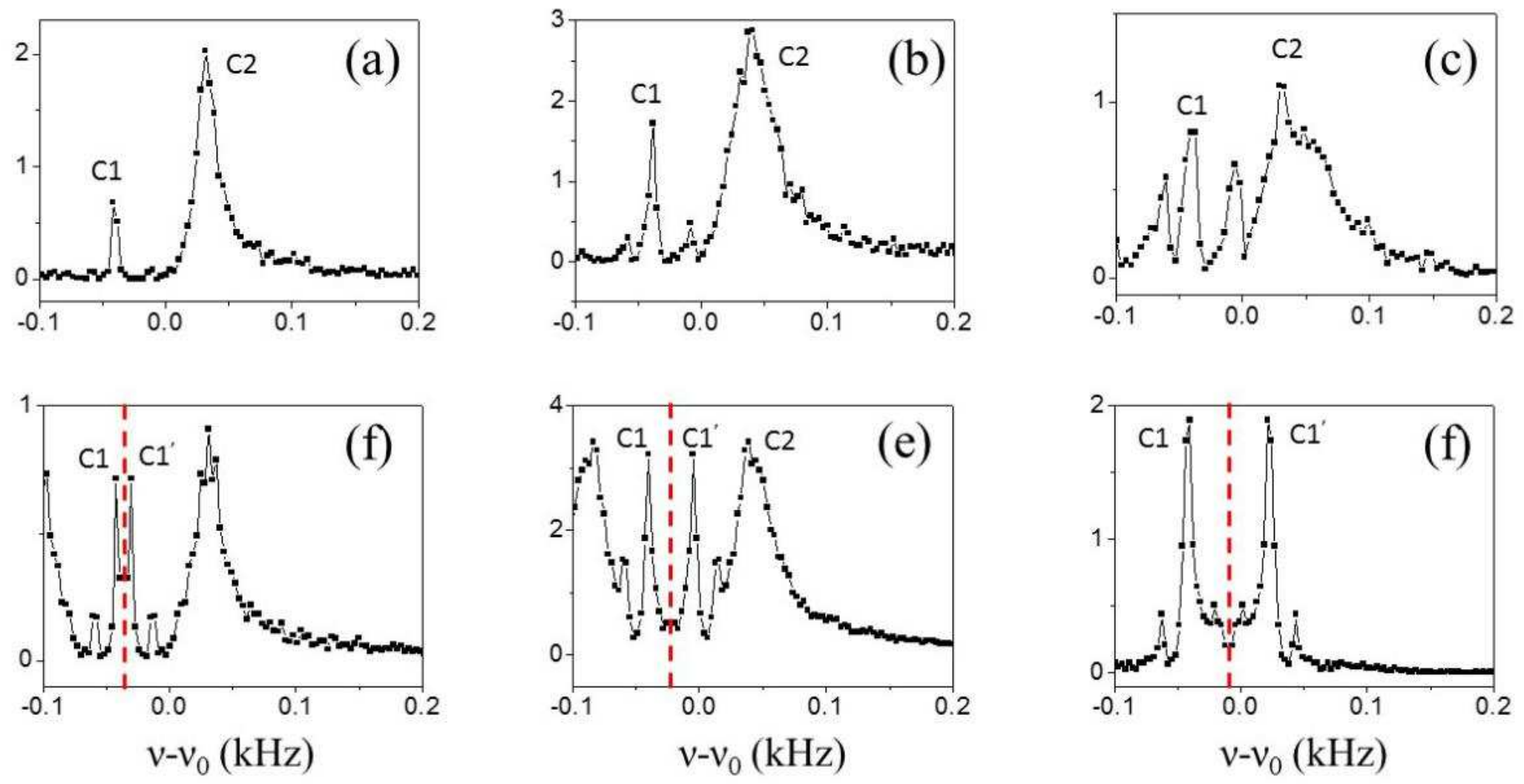

**Figure S8 | High-resolution spectroscopy of single nuclear spins (Data2)**. Fourier transform of the correlations constructed from Data2. In each measurement cycle the electron spin is repetitively read out for 40 times in (a-c) and 80 times in (d-f). The number of DD pulses is (a/d) $N_{\mathrm{p}} = 100$, (b/e) $N_{\mathrm{p}} = 200$, and (c/f) $N_{\mathrm{p}} = 300$. The red line at $2\pi\nu t_{\mathrm{c}} = 0 \bmod 2\pi$ is the symmetric line. The number of time points for Fourier transform is $N_{FT} = 1000$.

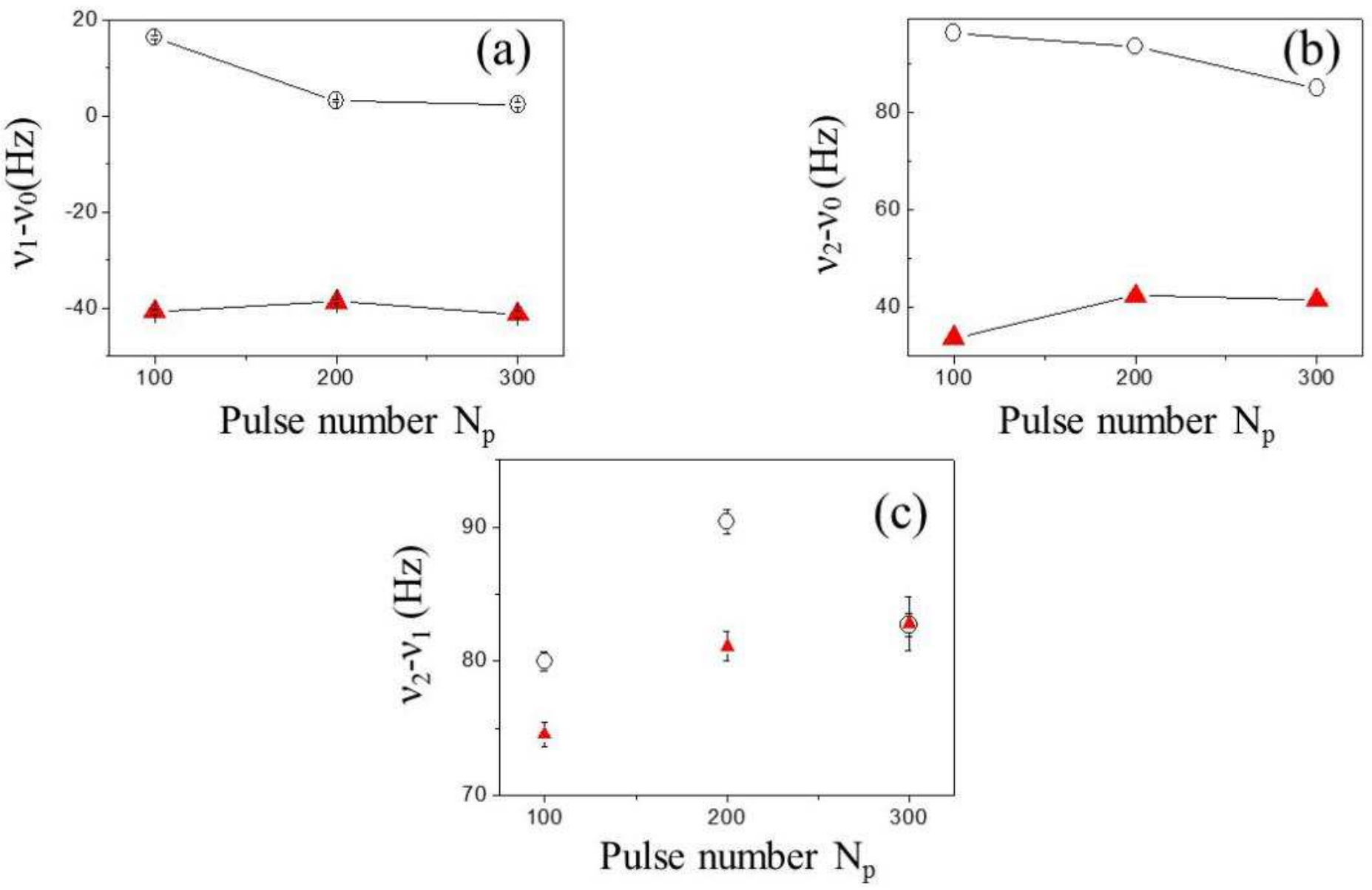

**Figure S9 | Shift of the bare Larmor frequency between the two data sets**. (a) Frequency of C1, (b) frequency of C2, and (c) frequency difference between C1 and C2, obtained from Data1 (black circles) and Data2 (red triangles), as functions of the DD pulse number.

# I. $^{14}$N nuclear spin assisted readout of NV centre electron spin

When detecting spins via the autocorrelation of subsequent measurements, the readout efficiency of the electron spin is a critical parameter. It can be increased by transferring the electron spin state to the $^{14}$N nuclear spin, which can be read out repetitively in a non-demolition way[2,3]. Since only the $m_S = 0$ and $m_S = -1$ spin manifold is used, the $^{14}$N nuclear spin needs to be initialized into a sub-manifold consisting of two eigenstates, in our case $m_I = 0$ and $m_I = +1$. This is done by two conditional $\pi$ rotation of the electron and nuclear spin, which constitutes a SWAP gate between the NV electron spin and the $^{14}$N nuclear spin. Subsequently, the NV electron spin is reinitialized (see Fig. S1 (a)). Due to insufficient initialization of the combined NV electron spin and charge state (around 70% $NV^-$ and 30% $NV^0$)[4], the procedure is done twice.

After the measurement, the electron spin state is again transferred onto the nuclear spin (see Fig. S1 (b)) by a SWAP gate. Owing to the stability of the $^{14}$N spin state during optical excitation, the nuclear spin state can be transferred repetitively to the electron spin state, and read out (40 times in our experiments). This method is used for Fig. 4 in the main text.

# II. Correlation of sequential weak measurement - formalism

## *II.1. Control phase gate by dynamical decoupling sequence*

Each measurement cycle includes four steps (see Fig. 1g in the main text):

1. the electron spin is initialized to the $x$ state (by initialization to the $z$ direction state and then rotation by a $(-\pi/2)_y$ pulse);
2. evolution under DD control for a period of $t_I$;
3. free evolution for a waiting time $t_{read}$;
4. Measurement of the electron spin $y$ component (realized by a rotation by a $(\pi/2)_x$ pulse and then measurement of the $z$ component via photoluminescence).

The NV center spin $\hat{\mathbf{S}}$ and the nuclear spin-1/2 $\hat{\mathbf{I}}$ have the interaction

$$\hat{H} = |+\rangle\langle+|\otimes\mathbf{v}_+\cdot\hat{\mathbf{I}} + |-\rangle\langle-|\otimes\mathbf{v}_-\cdot\hat{\mathbf{I}},$$

where $|+\rangle \equiv |m_S = 0\rangle$, $|-\rangle \equiv |m_S = -1\rangle$, $\mathbf{v}_+ \equiv \nu_0\mathbf{e}_z$, and $\mathbf{v}_- \equiv \nu_0\mathbf{e}_z + \mathbf{A}$. The evolution under the DD control is

$$\hat{U}_{\text{DD}} = \left[e^{-i2\pi\hat{H}\tau}e^{-i\pi\hat{S}_x}e^{-i2\pi\hat{H}\tau}\right]^{N_p},$$

where $\tau = t_I / N_p$. $\hat{U}_{\text{DD}}$ can always be factorized to

$$\hat{U}_{\text{DD}} = e^{-i\Phi\mathbf{n}\cdot\hat{\mathbf{I}}}e^{i2\alpha\hat{S}_z\hat{\mathbf{I}}_{\hat{\mathbf{a}}}}.$$

## *II.2. First-order Magnus expansion for weakly coupled nuclear spins*

We write the evolution as $\hat{U}_{\text{DD}} = \hat{U}_+ |+\rangle\langle+| + \hat{U}_- |-\rangle\langle-|$ with

$$\begin{aligned}\hat{U}_+ &= (e^{-i\tau\pi\mathbf{v}_-\cdot\hat{\mathbf{I}}}e^{-i\tau\pi\mathbf{v}_+\hat{I}_z})^{N_p},\\ \hat{U}_- &= (e^{-i\tau\pi\mathbf{v}_+\cdot\hat{\mathbf{I}}}e^{-i\tau\pi\mathbf{v}_-\hat{I}_z})^{N_p}.\end{aligned}$$

Here, we consider the case of even pulse numbers. $\hat{U}_\pm$ can be expressed as a time-ordered integration

$$\hat{U}_\pm = \mathrm{T}e^{-i2\pi\int_0^{t_I}\hat{H}_\pm(t)\mathrm{d}t},$$

with

$$\hat{H}_{\pm}(t) = \hat{\bar{H}} \pm \frac{1}{2}\beta(t)\mathbf{A}\cdot\hat{\mathbf{I}},$$

where $\hat{\bar{H}} = \bar{\mathbf{v}}\cdot\hat{\mathbf{I}}$, $\bar{\mathbf{v}} = \nu_0\mathbf{e}_Z + \mathbf{A}/2$, and the modulation function $\beta(t)$ alternates between +1 and -1 every time a π-pulse is applied.

In the interaction picture defined by $\hat{\bar{H}}$, the evolution operator is

$$\hat{U}_{\text{DD}} = e^{-i2\pi\bar{\mathbf{v}}\cdot\hat{\mathbf{I}}t_I}\mathrm{T}e^{-i2\pi\hat{S}_z\int_0^{t_I}\beta(t)\mathbf{A}\cdot\hat{\mathbf{I}}(t)\mathrm{d}t},$$

where $\hat{\mathbf{I}}(t) = e^{i2\pi\bar{\mathbf{v}}\cdot\hat{\mathbf{I}}t_I}\hat{\mathbf{I}}e^{-i2\pi\bar{\mathbf{v}}\cdot\hat{\mathbf{I}}t_I}$. We decompose $\mathbf{A}$ to $\mathbf{A} = \mathbf{A}_{\perp} + \mathbf{A}_Z$. The Z component part is averaged out and hence we obtain

$$U_{\text{DD}} = e^{-i2\pi\bar{\mathbf{v}}\cdot\hat{\mathbf{I}}t_I}e^{-i2\pi\hat{S}_z\int_0^{t_I}\beta(t)\mathbf{A}_{\perp}\cdot\hat{\mathbf{I}}(t)\mathrm{d}t}.$$

The first order Magnus expansion, which is valid for weak hyperfine interaction, gives[5]

$$\mathrm{T}e^{-i2\pi\hat{S}_z\int_0^{t_I}\beta(t)\mathbf{A}_{\perp}\cdot\hat{\mathbf{I}}(t)\mathrm{d}t} \approx e^{-i2\pi\hat{S}_z\int_0^{t_I}\beta(t)\mathbf{A}_{\perp}\cdot\hat{\mathbf{I}}(t)\mathrm{d}t} \equiv e^{2i\alpha\hat{S}_z\mathbf{e}_X\cdot\hat{\mathbf{I}}},$$

where

$$\alpha = \frac{2A_{\perp}}{\bar{\nu}}\frac{\sin\left(N_{\text{p}}\pi\bar{\nu}\tau\right)}{\cos\left(\pi\bar{\nu}\tau\right)}\sin^2\frac{\pi\bar{\nu}\tau}{2},$$

and $\mathbf{e}_X$ is the unit vector rotating from $\mathbf{A}_{\perp}$ by an angle $-\pi\bar{\nu}t_I$ around the $Z$ axis.

The readout period between the DD controls can be adjusted for fine tuning the nuclear spin precession. During the readout period, the perpendicular component of the hyperfine interaction $\mathbf{A}_{\perp}$ is averaged to zero by the fast precession of the nuclear spin (the Zeeman frequency ~MHz is much greater than the hyperfine interaction ~kHz in the experiment). The evolution of the nuclear spin during the readout period is $\hat{U}_{\text{read}} = \exp\left(-i2\pi\bar{\nu}t_{\text{read}}\hat{I}_Z\right)$ plus a small dephasing along the *Z* axis (which results from the random population of the electron spin states, see Sec. IV below for more discussions). Since the dephasing about the *Z* axis is small and does not affect the frequency of the nuclear spin precession in the *X*-*Y* plane, we can drop it in calculating the correlation function.

Therefore, the evolution of the electron spin and the nuclear spin, during the DD control and the readout period, can be written as

$$\hat{U}_T = e^{-i\Phi\hat{I}_Z} e^{2i\alpha\hat{S}_z\hat{I}_X},$$

a control phase gate plus a free precession by an angle $\Phi = 2\pi\bar{\nu}t_c$ with hyperfine-modified frequency $\bar{\nu}$.

## III. Correlation functions for multiple nuclear spins

For multiple nuclear spins, the evolution operator is $\hat{U}_\pm = \prod_i \hat{U}_{i,\pm}$ if we neglect the interaction between the nuclear spins. Here $\hat{U}_{i,\pm} = e^{-i\Phi_i\hat{I}_{i,Z}} e^{\pm i\alpha_i\hat{I}_{i,X}}$ is the unitary evolution operator of the *i*-th nuclear spin for the electron spin in the state $|\pm\rangle$. The Kraus operators of the multi-spin systems are $\hat{M}_\pm = \left(\prod_i e^{i\alpha_i\hat{I}_{i,X}} \pm i\prod_i e^{-i\alpha_i\hat{I}_{i,X}}\right)\Big/2$. The precession is $\hat{\mathcal{V}}[\hat{\rho}] = \left(\prod_i \hat{\mathcal{V}}_i\right)[\hat{\rho}]$ with $\hat{\mathcal{V}}_i[\hat{\rho}] \equiv e^{-i\Phi_i\hat{I}_{i,Z}}\hat{\rho}e^{i\Phi_i\hat{I}_{i,Z}}$. The correlation function is

$$C(N) \equiv \langle m_{k+N}m_k\rangle = \text{Tr}\left[\hat{\mathcal{P}}\left(\hat{\mathcal{V}}\hat{\mathcal{M}}\right)^{N-1}\hat{\mathcal{V}}\hat{\mathcal{P}}[\hat{\rho}_0]\right],$$

where the polarization operator $\hat{\mathcal{P}}[\hat{\rho}] \equiv \hat{M}_+\hat{\rho}\hat{M}_+^\dagger - \hat{M}_-\hat{\rho}\hat{M}_-^\dagger$ and $\hat{\mathcal{M}}[\hat{\rho}] \equiv \hat{M}_+ \hat{\rho} \hat{M}_+^\dagger + \hat{M}_- \hat{\rho} \hat{M}_-^\dagger$. The initial state $\hat{\rho}_0$ is taken as unpolarized.

In general the weak measurement of the multiple nuclear spins via the projective measurement of a commonly coupled central electron spin can introduce many-body correlations, which is interesting but will not be studied here. Instead, we consider the case that the measurement strength is weak, i.e., $|\alpha_i| << 1$. In this case, we have $\hat{\mathcal{P}}[\hat{\rho}] \approx \sum_i \alpha_i \hat{I}_{i,X}$ , which has no correlation in the leading order. Therefore, the following measurement induced dephasing and the precession can be considered as independent processes. The correlation function becomes a simple form as

$$C(N) \approx \sum_i C_i(N),$$

in which $C_i(N)$ is the correlation function of the sequential weak measurement if only the *i*-th nuclear spin is in the presence.

## IV. Dephasing during the readout

During the readout period (the measurement cycle excluding the DD control), the rapid Larmor precession of the nuclear spin averages the perpendicular hyperfine interaction to be zero (during the resonant DD control, on the contrary, the perpendicular component is preserved), and the effective coupling becomes

$$H_{\text{read}} = \bar{\nu}_0 \hat{I}_Z + \sum_j |j\rangle\langle j| \otimes A_j \hat{I}_Z,$$

where $|j\rangle$ denotes different levels (including the optically excited states and different charge states) of the NV centre and $A_j$ is the corresponding hyperfine constant. The second term in the r.h.s. of the equation above would induce dephasing of the nuclear spin along the $Z$ axis. Depending on the correlation time $\tau_{\mathrm{c}}$ of the electron spin random jumps, the dephasing during the readout time ($t_{\mathrm{read}}$) is estimated to be

$$\gamma' \approx \begin{cases} 2\pi^2 A_Z^2 t_{\mathrm{read}}{}^2 \text{ for } \tau_{\mathrm{c}} >> t_{\mathrm{read}} \text{(inhomogeneous broadening)}, \\ 4\pi^2 A_Z^2 t_{\mathrm{read}} \tau_{\mathrm{c}} \text{ for } \tau_{\mathrm{c}} << t_{\mathrm{read}} \text{ (motional narrowing)}, \end{cases}$$

and in between for intermediate correlation times. The dephasing rate can be accounted by an effective interaction time $\tau_{\mathrm{eff}}$ and the dephasing rate can be written as $A_Z^2 \tau_{\mathrm{eff}}^2$. In particular $\tau_{\mathrm{eff}} \approx 2\pi\sqrt{t_{\mathrm{read}}\tau_c}$ in the motional narrowing regime. From the extrinsic broadening of C2 peak in Fig.4 of main text, the effective interaction time is estimated to be 600 μs for the case that electron is repetitively readout for 40 times. Correspondingly, the dephasing in the $Z$ basis is $\gamma' = 1\times10^{-4}$, which is much smaller than $\gamma_{\mathrm{eff}} \approx 1\times10^{-3}$ for C1.

## V. Data Processing

### *V. 1.* Reconstruction of correlation function from photon counts

The correlation function is $C(N) = \sum_{m_k, m_{k+N}} m_k m_{k+N} p(m_k, m_{k+N})$, where $p(m_k, m_{k+N})$ is the joint probability for the two outputs under perfect projective measurements of the NV center. The correlation spectrum is obtained by Fourier transform $S(\nu) = \sum_{N=1}^{N_{\mathrm{FT}}} \exp(i2\pi N\nu t_{\mathrm{c}}) C(N)$.

#### *V.1.1. Direct reconstruction*

The raw data is the photon number $n_k$ collected from each measurement cycle. The joint probability of the two outcomes $n_k$ and $n_{k+N}$ depends on the correlation $C(N)$ through

$$p(n_k, n_{k+N}) = \sum_{m_k, m_{k+N}} D(n_k \mid m_k) D(n_{k+N} \mid m_{k+N}) p(m_k, m_{k+N}),$$

where $D(n_k \mid m_k)$ is the probability of detecting $n_k$ photons given that the electron spin is in the state that would produce the output $m_k$ under a perfect projective measurement . Using the relations $p(+,+) = p(-,-)$ and $p(+,-) = p(-,+)$, we obtain $p(m_k, m_{k+N}) = \sum_{m_k, m_{k+N}} \left[1 + m_k m_{k+N} C(N)\right]/4$ and hence the correlation of the photon counts

$$\begin{aligned}
\langle n_k n_{k+N} \rangle &= \sum_{n_k, n_{k+N}} n_k n_{k+N} p(n_k, n_{k+N}) \\
&= \frac{1}{4} \sum_{n_k, n_{k+N}} n_k n_{k+N} \sum_{m_k, m_{k+N}} D(n_k \mid m_k) D(n_{k+N} \mid m_{k+N}) \left[1 + m_k m_{k+N} C(N)\right] \\
&= \bar{n}^2 + \frac{1}{4} (\bar{n}_+ - \bar{n}_-)^2 C(N),
\end{aligned}$$

where $\bar{n}_\pm = \sum_n n D(n \mid \pm)$ is the average photon counts for the NV center in the $\pm$ state, and $\bar{n} = (\bar{n}_+ + \bar{n}_-)/2$ is the average photon count. Thus the correlation is constructed from the photon count statistics as

$$C(N) = \frac{4\left(\langle n_k n_{k+N} \rangle - \bar{n}^2\right)}{(\bar{n}_+ - \bar{n}_-)^2}.$$

Now we estimate the fluctuation of the correlation as reconstructed from the photon statistics, which is subjected to the shot noises. Since $\bar{n}^2$ is a constant, its fluctuation has no contribution to the correlation spectrum at nonzero frequencies.

The fluctuation of the denominator $(\bar{n}_+ - \bar{n}_-)^2$ would result in fluctuation of the overall amplitude of the correlation function and the spectrum without effects on the resonance frequency and width. Thus we only consider the fluctuation of the correlation $\langle n_k n_{k+N} \rangle$. For a finite sequence of $M$ outputs $\{n_1, n_2, \cdots, n_M\}$, the correlation is calculated by

$$\langle n_k n_{k+N} \rangle_M \approx \sum_{k=1}^{M-N} n_k n_{k+N} \Big/ (M-N).$$

For $N \ll M$, the fluctuation $\delta \langle n_k n_{k+N} \rangle_M \approx \delta n_k n_{k+N} / \sqrt{M}$ according to the center limit theorem, where $\delta n_k n_{k+N}$ is the fluctuation of two joint outputs. By direct calculation we obtain

$$\begin{aligned}\overline{(n_k n_{k+N})^2} &= \frac{1}{4} \sum_{n_k, n_{k+N}} (n_k n_{k+N})^2 \sum_{m_k, m_{k+N}} D(n_k \mid m_k) D(n_{k+N} \mid m_{k+N}) \left[1 + m_k m_{k+N} C(N)\right] \\ &= \frac{1}{4}\left(\overline{n_+^2} + \overline{n_-^2}\right)^2 + \frac{1}{4}\left(\overline{n_+^2} - \overline{n_-^2}\right)^2 C(N),\end{aligned}$$

where $\overline{n_\pm^2} \equiv \sum_{n_k} n_k^2 D(n_k \mid \pm)$, which is $\overline{n_\pm^2} = \bar{n}_\pm^2 + \bar{n}_\pm$ for Poisson distribution. Using the condition $C(N) \ll 1$, we obtain the fluctuation as

$$(\delta n_k n_{k+N})^2 = \overline{(n_k n_{k+N})^2} - \langle n_k n_{k+N} \rangle^2 = \left[\bar{n}^2 + \bar{n} + \frac{1}{4}(\bar{n}_+ - \bar{n}_-)^2\right]^2 - \bar{n}^4.$$

As a result, the noise amplitude becomes

$$\delta C(N) = \frac{\varepsilon_1}{\sqrt{M}}$$

with

$$\varepsilon_1 = 4\sqrt{\left[\bar{n}^2 + \bar{n} + (\bar{n}_+ - \bar{n}_-)^2/4\right]^2 - \bar{n}^4} \Big/ (\bar{n}_+ - \bar{n}_-)^2 .$$

### *V. 1.2 Reconstruction from single-shot readout*

When the NV center $^{14}$N nuclear spin is employed to assist readout of the NV center electron spin state (see Sec. I of SI), the photon counts in each cycle of measurement can be quite large ($|\bar{n}_+ - \bar{n}_-| >> 1$ ) and it is possible to choose a threshold photon count number $n_{\text{th}}$ between $n_-$ and $n_+$ such that the output is recorded as +1 or -1 if the photon counts in a cycle is above or below the threshold, i.e.,

$$s_k = \begin{cases} +1 & \text{for } n_k > n_{\text{th}}, \\ -1 & \text{for } n_k \leq n_{\text{th}}. \end{cases}$$

For the NV center electron spin $m_k = \pm$ state, the conditional probability of output $s_k = \pm 1$ is

$$\begin{aligned} p_+ &\equiv p(+1|+) = \sum_{n_k > n_{\text{th}}} D(n_k|+) = 1 - p(-1|+), \\ p_- &\equiv p(-1|-) = \sum_{n_k < n_{\text{th}}} D(n_k|-) = 1 - p(+1|-). \end{aligned}$$

When $p_+ = p_- = 1$, the photon detection constitutes a perfect single-shot measurement. The joint probability of two outputs is

$$p(s_k, s_{k+N}) = \sum_{m_k, m_{k+N}} p(s_k|m_k) p(s_{k+N}|m_{k+N}) p(m_k, m_{k+N}),$$

and correlation function is

$$\begin{aligned} \langle s_k s_{k+N} \rangle &= \sum_{s_k, s_{k+N}} s_k s_{k+N} p(s_k, s_{k+N}) \\ &= \sum_{s_k, s_{k+N}} s_k s_{k+N} \sum_{m_k, m_{k+N}} p(s_k|m_k) p(s_{k+N}|m_{k+N}) p(m_k, m_{k+N}) \\ &= \sum_{s_k, s_{k+N}} s_k s_{k+N} \sum_{m_k, m_{k+N}} p(s_k|m_k) p(s_{k+N}|m_{k+N}) \frac{1 + m_k m_{k+N} C(N)}{4} \\ &= (p_+ - p_-)^2 + (p_+ + p_- - 1)^2 C(N). \end{aligned}$$

Using $\langle s_k \rangle = \langle s_{k+N} \rangle = p_+ - p_-$, we reconstruct the correlation function of the weak measurements on the nuclear spin as

$$C(N) = \frac{\langle s_k s_{k+N}\rangle - \langle s_k\rangle\langle s_{k+N}\rangle}{(p_+ + p_- - 1)^2}.$$

With similar approaches in Sec.V.1, the correlation of the output shot noises is

$$\left(\delta s_k s_{k+N}\right)^2 = \left\langle s_k^2 s_{k+N}^2\right\rangle - \left\langle s_k s_{k+N}\right\rangle^2 \approx 1 - \left(p_+ - p_-\right)^4,$$

and the fluctuation of the correlation due to the photon shot noises

$$\delta C = \frac{\varepsilon_2}{\sqrt{M}},$$

with

$$\varepsilon_2 = \sqrt{1 - \left(p_+ - p_-\right)^4} \Big/ \left(p_+ + p_- - 1\right)^2.$$

## *V.1.3. Error bar of Fourier transformation*

We denote $N_{\mathrm{FT}}$ time-domain data by a real vector $\mathbf{s} = \{C(1), C(2), \ldots, C(N_{\mathrm{FT}})\}$. The Fourier transform is

$$\mathbf{f} = \mathbf{U}\mathbf{s}$$

where $U_{ij} = e^{-i2(i-1)(j-1)\pi/N_{\mathrm{FT}}}$. The matrix has the properties $\mathbf{U}^T = \mathbf{U}$ and $\mathbf{U}\mathbf{U}^\dagger = N_{\mathrm{FT}}$.

We assume that each elements of $\mathbf{s}$ has the normal distribution $\mathrm{N}\left(0, \sigma_t\right)$, where $\sigma_t$ denotes the noise amplitude in the time domain. We use $\mathbf{a} = \mathrm{Re}\mathbf{f}$ and $\mathbf{b} = \mathrm{Im}\mathbf{f}$ to denote the real part and imaginary part of $\mathbf{f}$. The covariance of $\mathbf{s}$ is

$$\langle \mathbf{s}\mathbf{s}^T\rangle = \sigma_t^2 \mathbf{1}.$$

The covariance of the Fourier transform are

$$\langle \mathbf{aa}^T \rangle = \left[ (\mathrm{Re}\mathbf{U}) \right] \langle \mathbf{ss}^T \rangle \left[ (\mathrm{Re}\mathbf{U}) \right]^T$$
$$\langle \mathbf{bb}^T \rangle = \left[ (\mathrm{Im}\mathbf{U}) \right] \langle \mathbf{ss}^T \rangle \left[ (\mathrm{Im}\mathbf{U}) \right]^T$$
$$\langle \mathbf{ab}^T \rangle = \left[ (\mathrm{Re}\mathbf{U}) \right] \langle \mathbf{ss}^T \rangle \left[ (\mathrm{Im}\mathbf{U}) \right]^T$$

After some simplification, the noise amplitude of real part and imaginary part is the same with each other and is linearly proportional to that of the time domain by the relation

$$\sigma_\omega = \sqrt{N_{\mathrm{FT}}/2}\sigma_t,$$

independent of the frequency.

When processing the experimental data, what we fit is the absolute value of the Fourier transform (not its real part and its imaginary part). As a result, we should give the noise amplitude of the absolute value of the Fourier transform. The fluctuation of the absolute value is

$$\delta S = \begin{cases} \sigma_\omega & S \gg \sigma_\omega \\ \sqrt{2 - \pi/2}\ \sigma_\omega & S \ll \sigma_\omega \end{cases}$$

In the following, we neglect the irrelevant constant (because it is in the order of 1), we obtain the spectrum fluctuation due to photon shot noise as

$$\delta S \approx \frac{\varepsilon_{1,2}}{\sqrt{M}} \sqrt{N_{\mathrm{FT}}/2}.$$

The spectrum height at the resonance frequency is

$$S(\nu_{\mathrm{eff}}) \approx \frac{1}{2} \min(N_{\mathrm{C}}, N_{\mathrm{FT}}) \sin^2 \alpha,$$

where $N_{\mathrm{C}} \sim \gamma_{\mathrm{eff}}^{-1}$ is the life time of the correlation signal. To maximize the signal-to-noise ratio, we choose $N_{\mathrm{FT}} \approx N_{\mathrm{C}}$ for constructing the correlation spectrum from the experimental data .

Figure S2 shows an example of the fluctuation of the correlation spectrum $|S(\nu)|$ (not its square), where the electron spin is repetitively read out for 40 times in each measurement cycle. The number of measurement cycles is about $M = 3\times10^{7}$. The averaged photon number collected per cycle is about $\bar{n} \approx 4.0$. The noise amplitude per cycle is $\varepsilon_1 \approx 28$. If we use $N_{\mathrm{FT}} = 1500$ time points for the Fourier transform, the noise amplitude of the spectrum is about $\delta S \approx 0.14$ and signal-to-noise ratio is about 10.6 for the highest peak as shown in Fig. S2.

## *V. 1.4. Comparison between the performance of the two methods*

We just need to compare the fluctuations of the output correlations $\varepsilon_1$ and $\varepsilon_2$ of the two methods in Sec. V1.1.2 & Sec. V.1.2. For the Poisson distribution of the photon counts $p(n|\pm) = e^{-\bar{n}_\pm}\bar{n}_\pm^n / n!$, the optimal threshold photon counts for single-shot readout is

$$n_{\mathrm{th}} = \left|\frac{\bar{n}_+ - \bar{n}_-}{\ln(\bar{n}_+/\bar{n}_-)}\right|,$$

which is between $\bar{n}_+$ and $\bar{n}_-$.

If the averaged photon number is very small ($\bar{n}_\pm \ll 1$), the threshold $n_{\mathrm{th}} \approx 0$, so $p_+ = \bar{n}_+$ and $p_- = 1 - \bar{n}_-$. The fluctuations in the output correlations in this low photon count limit are

$$\varepsilon_1 \approx \frac{1}{\eta^2 \bar{n}},$$

and

$$\varepsilon_2 \approx \frac{1}{\sqrt{2}\eta^2 \bar{n}^{3/2}},$$

where $\eta = (\bar{n}_+ - \bar{n}_-)/(\bar{n}_+ + \bar{n}_-)$ is the contrast of the fluorescence. The reconstruction from the single-shot readout method has much larger fluctuation in the low photon count limit.

In the large photon count limit ($\bar{n} >> 1$), $p_\pm \approx 1$. The fluctuations in the output correlations are

$$\varepsilon_1 \approx \sqrt{1 + 2\eta^{-2}},$$

and

$$\varepsilon_2 \approx 1.$$

The single-shot readout approach would be much better when the contrast is small.

Figure S3 (a) shows $\varepsilon_1$ (black line) and $\varepsilon_2$ (red line) as a function of $\bar{n}$ when the contrast is fixed to 0.18. As expected $\varepsilon_2$ is larger than $\varepsilon_1$ when $\bar{n} \lesssim 0.5$ and the opposite when $\bar{n} \gtrsim 0.5$. We also show the ratio $\varepsilon_2/\varepsilon_1$ as a function of $\bar{n}$ in Fig. S3 (b) for different contrasts. In all cases the single-shot readout scheme has larger fluctuation for $\bar{n} \lesssim 0.5$ and smaller fluctuations for $\bar{n} \gtrsim 0.5$. In our experiments, the largest $\bar{n}$ is about 4.0 when the electron state is repetitively read out via the auxiliary nitrogen nuclear spin (as for Fig. 4 in the main text). Under this condition, the single-shot readout approach has smaller fluctuations by a factor 2 than the direct reconstruction. Without repetitively readout, $\bar{n}$ is about 0.1 (as for Figs. 2 & 3 in the main text). Under such cases, the direct reconstruction approach has smaller fluctuations by a factor of 2.3 than the single-shot readout scheme. Since there is no orders of magnitude difference in fluctuations between the two methods, we use the direct reconstruction in all the cases.

## *V. 2.* Spatial sensing range

We consider the weak measurement limit for estimation of the sensing range.

The resonance signal of the correlation spectrum is

$$S(\nu_{\text{eff}}) \approx \frac{1}{2} N_{\text{C}} \sin^2 \alpha \approx 2 N_C A_\perp^2 t_{\text{I}}^2,$$

which is reached when the number of time points used for Fourier transform is $N_{\text{FT}} = N_{\text{C}}$. The correlation lifetime $N_{\text{C}} \approx 1/(\gamma_{\text{eff}} + \gamma_{\text{ex}})$ (in units of measurement cycles), with the measurement induced dephasing per cycle $\gamma_{\text{eff}} = A_\perp^2 t_I^2$ and the dephasing during the waiting and readout periods $\gamma_{\text{ex}} \approx A_Z^2 \tau_{\text{eff}}^2$. The noise amplitude in the frequency domain becomes

$$\delta S = \frac{\varepsilon_{1(2)}}{\sqrt{M}} \sqrt{N_{\text{C}}},$$

where the fluctuation $\varepsilon_{1(2)}$ depending on the reconstruction method. The signal-to-noise ratio is

$$\text{SNR} = \frac{\sin^2 \alpha}{2} \frac{\sqrt{M N_{\text{C}}}}{\varepsilon_{1(2)}} \approx \frac{2 A_\perp^2 t_{\text{I}}^2}{\sqrt{A_\perp^2 t_{\text{I}}^2 + A_Z^2 \tau_{\text{eff}}^2}} \frac{\sqrt{M}}{\varepsilon_{1(2)}}.$$

Using the position dependence of the hyperfine interaction $A_Z = A_0 \left(1 - 3\cos^2\theta\right)/d^3$ and $A_\perp = 3 A_0 \cos\theta \sin\theta / d^3$ (for $^{13}C$ nuclear spin $A_0 \approx -20\ \text{kHz} \cdot \text{nm}^3$), we obtain the relation between signal-noise ratio and the spatial position of the target nuclear spin

$$\mathrm{SNR} = \frac{d_0^3}{d^3}\left|\sin(2\theta)\right| \Big/ \sqrt{1+\left(\frac{1+3\cos(2\theta)}{3\sin(2\theta)}\frac{\tau_{eff}}{t_\mathrm{I}}\right)^2},$$

where $d$ and $\theta$ is the distance from the central spin and polar angle, and

$$d_0 = \left(\frac{3A_0 t_\mathrm{I}\sqrt{M}}{\varepsilon_{1(2)}}\right)^{1/3},$$

is the typical sensing distance. For the current parameters, the measurement times $M = 3\times10^7$ and $\varepsilon_1 = 28$ for 40 times repetitive readout. For the case of 100 pulse number ($t_I \approx 18\mu\mathrm{s}$), we estimate that the typical detecting distance is $d_0 \approx 6\mathrm{nm}$. Figure S4 shows some examples of the spatial range of sensing for various interaction times.

The ultimate sensing range is limited by the coherence time of the NV center ($t_\mathrm{I} \le T_2$) and the photon-shot noise $\varepsilon_{1(2)} \ge 1$. For an NV center in bulk diamond with coherence time ~ 1 ms, $d_0^{\max} = (3A_0T_2)^{1/3} M^{1/6} \approx 4M^{1/6}$ (nm). For a shallow NV center located about 8 nm below diamond surface, the coherence time can reach about 100 μs [6], the corresponding sensing range would be about 16 nm. For a hydrogen nuclear spin with distance 8 nm away from such an NV center, the coupling is about 0.15 kHz. The corresponding spectral resolution is estimated to be 0.3 Hz.

## *V. 3. Correction of systematic errors*

The systematic errors that affect the NV center fluorescence, including slow spatial drift of the NV center out of the microscope focus and oscillation of the laser output power due to the power grid will be reflected in the correlation function. We correct these effects by fitting and subtracting an exponential decay, which results from the slow focus drift, and also fit and subtract a 100 Hz oscillation of the form, which stems from a 100 Hz modulation of the diode laser output power, due to the rectified 50 Hz AC power grid.

## *V. 4. Determination of hyperfine interaction*

The hyperfine interaction $A_Z$ is determined by $\Phi = (\nu_0 + A_Z/2)t_c \bmod 2\pi$, which has ambiguity in multiples of $2\pi$. Further ambiguity is caused by the symmetry of the correlation function under the transformation $\Phi \leftrightarrow -\Phi \bmod 2\pi$.The first kind of ambiguity is removed as long as the hyperfine-renormalized Larmor frequency has been roughly determined with precision better than $1/t_c$ (e.g., by the resonant condition of DD control). The second kind of ambiguity can be removed by varying $t_c$, which shift the resonance peaks of $\Phi$ and $2\pi - \Phi$ toward opposite directions in frequency. The correlation spectrum $|S(\nu)|$ (not its square) is shown in Fig. S5 Among all the peaks around $\nu - \nu_0 \approx 0.572\text{kHz}$, the common peak for all values of $t_c$ (indicated by the red arrow) is the resonance peak, which yields $\bar{\nu} = \nu_0 + A_Z/2$ and hence $A_Z = 1.144$ kHz with error 54Hz.

## *V. 5. Determine the error bar of the parameters*

For a set of data $\{\nu_i, f_i\}, i = 1, \cdots, N_{\text{FT}}$ as the signal (for example, the signal in the frequency domain), we fit it by a theoretical function $f(\nu, \boldsymbol{\lambda})$ with parameters being $\boldsymbol{\lambda}$. We assume that the signal $f_i$ obeys a Gaussian distribution $\text{N}(0, \sigma_i)$. According to the Bayesian formula, the distribution of fitting parameters $\boldsymbol{\lambda}$ conditioned on the signal now becomes

$$P(\boldsymbol{\lambda} | \mathbf{x}) \propto \exp\{-\Delta(\boldsymbol{\lambda})\},$$

where

$$\Delta(\boldsymbol{\lambda}) = \sum_{i=1}^{N_{\text{FT}}} \frac{\left[f_i - f\left(v_i, \boldsymbol{\lambda}\right)\right]^2}{2\sigma_i^2}$$

is the cost function. The optimized estimation of $\boldsymbol{\lambda}$ is $\boldsymbol{\lambda}_e$ which maximizes the distribution $P(\boldsymbol{\lambda} | \mathbf{x})$, or minimize $\Delta(\boldsymbol{\lambda})$ with

$$\frac{\partial \Delta(\boldsymbol{\lambda})}{\partial \boldsymbol{\lambda}} |_{\boldsymbol{\lambda}=\boldsymbol{\lambda}_e} = 0.$$

Around the peak, the distribution can be expanded as

$$P(\boldsymbol{\lambda} | \mathbf{x}) \propto \exp\left\{-\frac{1}{2}(\boldsymbol{\lambda} - \boldsymbol{\lambda}_e)\boldsymbol{\Sigma}(\boldsymbol{\lambda} - \boldsymbol{\lambda}_e)\right\},$$

where

$$\boldsymbol{\Sigma} = \frac{\partial}{\partial \boldsymbol{\lambda}}\left(\frac{\partial \Delta(\boldsymbol{\lambda})}{\partial \boldsymbol{\lambda}}\right)\Bigg|_{\boldsymbol{\lambda}=\boldsymbol{\lambda}_e}$$

is a matrix. As a result, the covariance matrix

$$\langle(\boldsymbol{\lambda} - \boldsymbol{\lambda}_e)(\boldsymbol{\lambda} - \boldsymbol{\lambda}_e)\rangle = \boldsymbol{\Sigma}^{-1}$$

and the error bar of each parameters is

$$\delta\lambda_i = 2\sqrt{\left(\boldsymbol{\Sigma}^{-1}\right)_{ii}}.$$

As an example, we fit the peak in Fig.S5. Here, the fitting curve is

$$f\left(v, \boldsymbol{\lambda}\right) = A\sqrt{\frac{\gamma^2}{(x - x_0)^2 + \gamma^2}}$$

and the fitting parameters is $\boldsymbol{\lambda} = \{A, x_0, \gamma\}$ . $\sigma_i = \sqrt{N_{\text{FT}} / (2M)}$ is the noise amplitude in the frequency domain. For the XY8 case, $M = 1.9\times10^{10}$. We choose $N_{\text{FT}} = 2000$ and hence $\sigma_i = 0.08$. Using the above theory, the parameter is fitted to $x_0 = 0.572\text{kHz} \pm 27\text{Hz}$ and $\gamma = 73\text{Hz} \pm 47\text{Hz}$, where $x_0$ is the estimation of the difference $\bar{v} - v_0$ between the precession frequency and the bare Larmor frequency. The error bars in the figures of the main text are obtained similarly.

## *V. 6. High-precision measurement of the bare Larmor frequency*

To determine the strength of the longitude hyperfine coupling, the bare Larmor frequency of the $^{13}C$ nuclear spins is determined with high precision using the method in Ref. [1]. Since this measurement relies on a correlation spectroscopy scheme, where free evolution of the detected spins occurs, while the NV electron spin is initialized into the $m_S = 0$ state, no hyperfine coupling is visible in the resulting spectrum. The method measures the difference between the Larmor frequency and the frequency of an inductive radio-frequency wave with $\nu_{\text{in}} = 2.730133\ \text{MHz}$. The difference is measured to be $\nu_0 - \nu_{\text{in}} = 10.0005\ \text{kHz} \pm 0.39\ \text{Hz}$ (Fig. S6). As a result, the bare Larmor frequency is $\nu_0 = 2.740134\ \text{MHz} \pm 0.39\ \text{Hz}$.

## *V. 7. High resolution spectroscopy by NV2*

The high resolution spectroscopy in Fig. 4 of the main text is performed on NV2 instead of NV1 (which is measured for Figs. 2 & 3 of the main text). The measurement induced back action on the target nuclear spin is very small because the target is relatively far away from the central spin. This experiment has been carried out in two runs, generating two sets of data, Data1 and Data2. The electron spin is read out repeatedly for 40 times in Data1 and both for 40 times and 80 times in Data2. The fluctuations of the correlation due to the photon shot noises are estimated to be $\varepsilon_1 = 28$ and $\varepsilon_1 = 20$ for case that electron is repetitively readout for 40 times and 80 times, respectively.

For Data1, the bare Larmor frequency of the nuclear spins has been measured very precisely with the method in Ref. [1]. The correlation spectrum is shown in Fig. S7. In Fig. S7(a), two nuclear spins are resolved. As the measurement strength is increased by applying more DD pulses, the peak width is broadened and more nuclear spins are resolved. In Fig. S7(c), four nuclear spins are detected. We concentrate on studying the C1 and C2 nuclear spins since these two nuclear spins are detected in all the three cases of Fig. S7.

The correlation spectrum from Data2 is shown in Fig. S8. Figure S8 (a-c) presents the correlation spectrum when electron state is repetitively read out for 40 times. The peaks in Fig.S7 are also found in Fig. S8. Figure S8 (d-f) presents the correlation spectrum when the electron state is repetitively read out for 80 times. In Fig. S8 (f), the C2 peak disappears, which is ascribed to the coherent trapping effect. For $t_c = 322.62400$ μs as chosen in this case, $|\sin\Phi| \approx 0.06$ for peak C2 is very close to $\tan^2(\alpha/2) \approx 0.05$ (for the estimated value $A_\perp \approx 4.02$ kHz in the main text). As there is some uncertainty in estimating $A_\perp$, the real value of $|\sin\Phi|$ may be smaller than $\tan^2(\alpha/2)$. As a result, the spin C2 is coherently trapped and its resonance is pinned at zero frequency (relative to $\nu_0$), which is not observed in the spectra since the static background has been subtracted from the correlation function in our data processing.

In comparison with the spectra from Data1, the peaks in the spectra from Data2 are shifted overall by about 40Hz. This shift comes from the shift of the magnetic field because between the two runs of experiments the crystal was moved inside the slightly inhomogeneous magnetic field. In Fig. S9 (a) and (b), we plot the peak positions of C1 and C2 for the two data sets. The shifts of C1 and C2 are both around 40Hz. We also plot the frequency difference between C1 and C2 for these two data sets. The difference is nearly unchanged both for different pulse numbers and different data sets. From the shift of the resonances from Data1 to Data2, the Larmor frequency for the Data2 is calibrated to $\nu_0' = 2.740090\ \text{MHz} \pm 7.9\ \text{Hz}$. The calibrated Larmor frequency is used in the correlation spectra shown in Fig. 4 of the main text.

## VI. Data acquisition time

We neglect the background decoherence of the nuclear spins. Under this condition, the correlation of sequential weak measurements has the form

$$C(N) = 8\pi\Delta\nu t_c e^{-N2\pi\Delta\nu t_c} \cos 2N\pi\nu_{\text{eff}} t_c,$$

where $\Delta\nu = \sin^2\alpha/(8\pi t_c)$ is the resolution of the frequency and $\alpha$ quantifies the measurement strength. If we use $N_{\text{FT}}$ data points to perform the Fourier transform, the optimized peak signal saturates to its maximum

$$S(\nu_{\text{eff}}) = 2,$$

when $N_{\text{FT}} \approx 1/(2\pi\Delta\nu t_c)$. The noise amplitude for the spectrum is

$$\delta S = \varepsilon_{1,2}\sqrt{\frac{N_{\text{FT}}}{2M}} = \varepsilon_{1,2}\sqrt{\frac{1}{4\pi\Delta\nu T^{\text{D}}}},$$

where the data acquisition time $T^{\text{D}} = Mt_c$. For a given signal-to-noise ratio $\text{SNR} \equiv S(\nu_{\text{eff}})/\delta S$ the data acquisition time to achieve a given resolution is thus

$$T^{\text{D}} = \text{SNR}^2 \frac{\varepsilon_{1,2}^2}{16\pi\Delta\nu}.$$

***Data acquisition time for the Ramsey scheme***

The Ramsey protocol begins with a measurement on the target though a DD sequence (with duration $t_I$) and electron spin measurement, then a free precession time is inserted, and finally another measurement is implemented. The precession time $t$ is swept from 0 to $T$ with a step $\tau$. For each precession time, the protocol is repeated for $M$ times to obtain the correlation between the two measurements. The final signal is the Fourier transform of the correlation.

Since there is no measurement in the precession time, there is no back-action and hence the correlation signal has the form

$$C(t) = 4\sin^2 \alpha_R \cos 2\pi\nu_{\text{eff}} t$$

$\alpha_R = 2A_{\perp} t_I$ quantifies the measurement strength in the DD process. In the following, we estimate the data acquisition time for achieving a given resolution.

The discrete Fourier transform gives

$$S(\nu_j) = 4\sin^2 \alpha_R \sum_{n=1}^{T/\tau} \cos(2\pi n \nu_{\text{eff}} \tau) e^{2\pi i n \nu_j \tau} = 2\sin^2 \alpha_R \left[ \frac{1-e^{i2\pi(\nu_j+\nu_{\text{eff}})T}}{1-e^{i2\pi(\nu_j+\nu_{\text{eff}})\tau}} + \frac{1-e^{-i2\pi(\nu_j-\nu_{\text{eff}})T}}{1-e^{i2\pi(\nu_j-\nu_{\text{eff}})\tau}} \right]$$

,

where $\nu_j = (j-1)/T$. The peak signal of the FFT is given by

$$S(\nu_{\text{eff}}) \approx 2\frac{T}{\tau}\sin^2 \alpha_R .$$

The noise amplitude is

$$\delta S = \frac{\varepsilon_{1,2}}{\sqrt{M}}\sqrt{\frac{T}{2\tau}} ,$$

with $\varepsilon_{1,2}/\sqrt{M}$ giving the noise amplitude of each data in the time domain. The signal-to-noise ratio becomes

$$\text{SNR} = \frac{S(\nu_{\text{eff}})}{\delta S} = \frac{4\sin^2 \alpha_R}{\varepsilon_{1,2}}\sqrt{\frac{T^D}{T}} = \frac{4\sin^2 \alpha_R}{\varepsilon_{1,2}}\sqrt{2\pi T^{\text{D}} \Delta \nu}$$

where $T^{\mathrm{D}} = MT^2/(2\tau)$ is the total data acquisition time and $\Delta\nu = 1/(2\pi T)$ is the spectrum resolution.

As a result, the data acquisition time for achieving a given resolution $\Delta\nu$ and signal-to-noise ratio SNR becomes

$$T^{\mathrm{D}} = \frac{1}{2\sin^4\alpha_R}\left(\mathrm{SNR}^2\frac{\varepsilon_{1,2}^2}{16\pi\Delta\nu}\right),$$

which has an extra factor $1/(2\sin^4\alpha_R)$ in comparison with the data acquisition time for the sequential weak measurement protocol.

Since no measurement is performed in the precessing process, the measurement back action on the target nuclear spin is absence. Hence, one can maximize the measurement strength by choosing proper DD duration $t_I$. The final result is

$$T^{\mathrm{D}} = \left(\mathrm{SNR}^2\frac{\varepsilon_{1,2}^2}{16\pi\Delta\nu}\right)\times\begin{cases}\frac{1}{2}, \text{if } A_\perp > A_c, \\ \frac{1}{2}\left(\frac{A_c}{A_\perp}\right)^4, \text{if } A_\perp < A_c,\end{cases}$$

where $A_c = 1/(2T_1)$ and $T_1$ is the life time of electron spin. For sensing a weakly coupled nuclear spin, the Ramsey protocol requires a data acquisition time longer by a factor of $\sim 1/(2A_\perp T_1)^4$ than the sequential weak measurement method.

## Supplementary References